\documentclass[aps,prl,onecolumn,noshowpacs,superscriptaddress,groupedaddress]{revtex4}  % for review and submission
\usepackage{epsfig,dsfont,amssymb,amsmath,amsthm,amsfonts,amsbsy,mathrsfs} 
\usepackage{lipsum}
\usepackage{color}
\usepackage{graphicx} 
\usepackage{amsmath} 
\usepackage{amssymb}
\usepackage{subfigure}
\usepackage{float}
\usepackage{hyperref} 
\usepackage{verbatim}
\usepackage{siunitx}
\usepackage[font=small,format=plain,labelfont=bf,justification=raggedright,singlelinecheck=false]{caption}
\newcommand {\bm}{\boldsymbol}

\begin{document}

\title{Cell Motility Dependence on Adhesive Wetting}

\author{Yuansheng Cao}
\thanks{These two authors contributed equally.}
\affiliation{Department of Physics, University of California, San Diego, La Jolla, California 92093, USA}
\author{Richa Karmakar}
\thanks{These two authors contributed equally.}
\affiliation{Department of Physics, University of California, San Diego, La Jolla, California 92093, USA}
\author{Elisabeth Ghabache}
\affiliation{Department of Physics, University of California, San Diego, La Jolla, California 92093, USA}
\author{Edgar  Gutierrez}
\affiliation{Department of Physics, University of California, San Diego, La Jolla, California 92093, USA}
\author{Yanxiang Zhao}
\affiliation{Department of Mathematics, The George Washington University, Washington, DC 20052, USA}
\author{Alex Groisman}
\affiliation{Department of Physics, University of California, San Diego, La Jolla, California 92093, USA}
\author{Herbert Levine}
\affiliation{Department of Bioengineering, Center for Theoretical Biological Physics, Rice University, Houston, Texas 77005, USA}
\author{Brian A. Camley}
\affiliation{Department of Physics \& Astronomy and
	Department of Biophysics, Johns Hopkins University, Baltimore, Maryland 21218, USA}
\author{Wouter-Jan Rappel}
\email{rappel@physics.ucsd.edu}
\affiliation{Department of Physics, University of California, San Diego, La Jolla, California 92093, USA}

\begin{abstract}
Adhesive cell-substrate interactions are crucial for cell motility and are
responsible for the necessary traction that propels cells. These interactions can also change the shape of the cell, analogous to liquid droplet wetting on adhesive substrates. To address how these shape changes affect cell migration and cell speed we model motility using deformable, 2D cross-sections of cells  in which
adhesion and frictional forces between cell and substrate can be
varied separately. 
Our simulations show that increasing the adhesion results in increased spreading of cells and larger cell speeds. We propose an analytical model which shows that the cell speed is inversely proportional to an effective height of the cell and that increasing this height results in increased internal shear stress. The numerical and analytical results are confirmed in experiments on motile eukaryotic cells. 
\end{abstract}

\maketitle

\section{Introduction}
Migration of eukaryotic cells plays an important role in many biological processes including development\cite{munjal2014}, chemotaxis\cite{kolsch2008}, and cancer invasion\cite{wirtz2011}. Cell migration is a complex process, involving external cues, intra-cellular biochemical pathways, and 
force generation. The adhesive interaction between cells and their extracellular environment is an essential part of cell motility \cite{geiger2009} and is generally thought to be responsible for frictional forces necessary for propulsion \cite{charras2014,chan2008traction}.
These frictional forces are due to the motion of the cytoskeleton network and
can be measured by 
traction force microscopy \cite{harris1980silicone}. 
On the other hand, adhesive cell-substrate interaction can 
also lead to cell spreading in both moving and non-moving cells \cite{frisch2002,keren2008, reinhart2005}. 
This is similar to the spreading of a liquid droplet during the wetting
of  an adhesive substrate.  
The resulting changes of the cell shape 
can potentially affect cell motility.
Experimentally, it is not possible to decouple the effect of 
adhesion and friction,  making it challenging to quantify the relative importance of spreading in cell motility. 

Here we investigate the dependence of motility on cell-substrate adhesion 
using a mathematical model in which we can alter the 
adhesive forces independent of frictional forces. We carry out numerical simulations of this model using the phase field approach, ideally suited 
for objects with deforming free boundaries \cite{kockelkoren2003, ziebert2013}. 
We  focus on a 2D vertical cross-section of a migrating cell which captures both 
cell-substrate interactions and internal fluid dynamics \cite{tjhung2015minimal}.  
 Our adhesive interactions are based on the phase-field description of wetting \cite{zhao2010,mickel2011} and are independent of the  molecular details of cell-substrate adhesion.
 Our simulations, together with an analytical 2D model extended from a previous 1D model \cite{carlsson2011}, generate several
  nontrivial and testable predictions which are
subsequently verified by  experiments using motile \text{\it Dictyostelium discoideum} cells.

\begin{figure}
\centering
\includegraphics[width=0.45\textwidth]{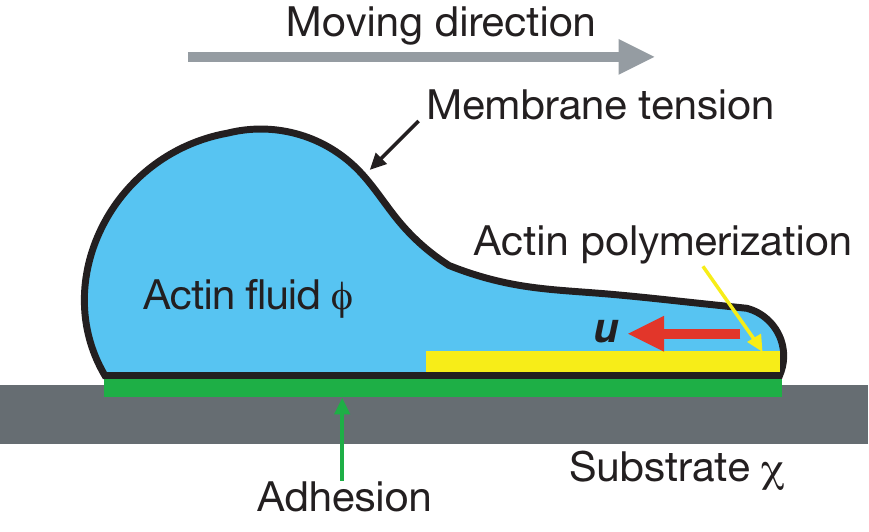}
\caption{Schematic illustration of a model cell on 
	a substrate. The cross-section of the cell is represented by 
	a phase field $\phi$ while the substrate is defined by a field $\chi$.
	The dynamics of the cytoskeleton network is modeled as an actin fluid with 
	velocity $\bf{u}$ (red arrow). Forces in the model include the membrane tension,  cell-substrate adhesive forces, forces due to active actin polymerization, and cytosolic viscous forces (proportional to $\bf{u}$). Actin polymerization is restricted to
	a narrow region near the substrate at the cell front-half, as indicated by the yellow band. Additional model details are given in main text and 
	in Supplemental Material.}
\label{fig1}
\end{figure}

\section{Results}
\subsection{Model}
Our vertical cross-sectional model cell captures the interaction 
of the cell with  the bottom  and, possibly, 
top substrate,  as well as the interior of the cell \cite{tjhung2015minimal} (Fig. \ref{fig1}).
This is in contrast to most computational studies of 
cell motility which model a flat cell that is entirely in contact with the substrate\cite{stephanou2008,mogilner2009mathematics,Bueetal10, alonso2018modeling}.  
This interior consists of a viscous cytoskeleton and is described as a compressible actin fluid \cite{rubinstein2009} with constant viscosity while cell movement is driven by active stress, located at the front of the cell. Note that we do not consider myosin-based contraction. 
Furthermore, and  following Ref. \cite{rubinstein2009}, we neglect the coupling between the actin fluid (representing the cytoskeleton) and the cytoplasm. 
	The latter is assumed to be incompressible, resulting in volume conservation.
	This type of model which treats the cytoskeleton as an active
	viscous compressible fluid  has been used in several recent studies 
	\cite{barnhart2011, bois2011pattern, shao2012, camley2013, goff2017actomyosin}.
Friction is caused by the motion of the 
cytoskeleton relative to the substrate and 
is taken to be proportional to the actin fluid velocity.  
To accurately capture cell shape and its deformations, we use the phase 
field approach in which an auxiliary  field $\phi(\boldsymbol{r},t)$ is introduced to distinguish between the interior ($\phi=1$) and exterior ($\phi=0$). This approach  allows us to efficiently track the cell boundary 
which is determined by  $\phi(\bm{r},t)=1/2$ \cite{shao2010,shao2012,camley2013,najem2013phase,ziebert2013,alonso2018modeling}.
In our model, boundary motion is driven by fluid flow which is determined by adhesion, friction, membrane forces and active protrusion. The cell is placed on a substrate which is parallel to the  $x$ direction, and polarized in one direction. 
As described in earlier work \cite{biben2005phase,shao2012,camley2013}, the evolution of the cell's shape is determined by the phase field dynamics:
\begin{equation}\label{phase_field}
\frac{\partial\phi(\bm{r},t)}{\partial t}=-\bm{u}\cdot\nabla\phi(\bm{r},t)+\Gamma(\epsilon\nabla^2\phi-G^{\prime}/\epsilon+c\epsilon|\nabla\phi|),
\end{equation}
where the advection term couples  the velocity field  of the actin fluid, $\bm{u}$, to the phase field,  $\epsilon$ is the width of the boundary, $\Gamma$ is a relaxation coefficient, $G$ is a double-well potential with minima 
at $\phi=1$ and $\phi=0$,   and $c$ is the local curvature of the boundary
(see Supplemental Material).

The actin fluid
velocity field is determined by the stationary Stokes equation with an
assumption of perfect compressibility  (zero pressure and neglecting the inertial term because of low Reynolds number) \cite{rubinstein2009,bois2011}:
\begin{equation}\label{Stokes}
\nabla\cdot[\nu \phi(\nabla\bm{u}+\nabla\bm{u}^T)]+\bm{F}_{sub}+\bm{F}_{mem}+\bm{F}_{area}+\nabla\cdot\sigma^a=0,
\end{equation}
where $\nu$ is the viscosity of the cell and where 
$\sigma^a$ is the active stress due to actin polymerization, 
further detailed below. 
$\bm{F}_{sub}$ is the interaction between the cell and substrate and contains both adhesion and friction, $\bm{F}_{sub}=\bm{F}_{adh}+\bm{F}_{fric}$. The adhesive force is given by $\bm{F}_{adh}=\frac{\delta H(\phi,\chi)}{\delta \phi}\nabla\phi$, with the cell-substrate interaction potential:
\[
H(\phi,\chi)=\int d\bm{r}^2\phi^2(\phi-2)^2 W(\chi).
\]
Here, $\chi(\bm{r})$ is a constant  field which marks the substrate (or ceiling) and continuously changes from $\chi=1$ (within the substrate) to $\chi=0$ (out of substrate;
Fig. S1).  $W(\chi)$ is a potential with a negative adhesion energy per unit length controlled by a parameter $A$ such that larger values of $A$ represent
a larger adhesive force between cell and substrate. In addition, 
this potential contains a short-range repulsion 
that ensures that the cell does not penetrate the substrate.
The term $ \phi^2(\phi-2)^2$ is added to ensure that the force peaks within 
the boundary and vanishes at $\phi=0$ and $\phi=1$.

The second term in $\bm{F}_{sub}$ describes the  frictional force between the cell and the substrate. 
Depending on the cell type, these forces can arise from 
focal adhesions  or from non-specific cell-substrate interactions.
For simplicity, the frictional force  in our cross-sectional model 
is modeled as a  
viscous drag proportional to the actin fluid velocity :
\[
\bm{F}_{fric}=-\xi_s\chi\bm{u} -\xi_d\bm{u},
\]
where the first term is the cell-substrate friction, parameterized 
by the coefficient $\xi_s$, and
the second term represents a damping force, introduced to increase numerical stability. 
We have verified that  the cell speed changes little when we
vary the drag coefficient $\xi_d$ (Fig. S2).
 Initially, we will vary both the adhesion energy (which controls spreading) and the frictional drag separately, allowing us to determine its 
    relative contribution to cell motility.
   We will then examine model extensions   which implement dependent adhesion and friction 
    	mechanisms  (see Fig. 4).  The uniform membrane tension $\bm{F}_{mem}$ and a force arising from cell area conservation $\bm{F}_{area}$ are introduced as 
in our previous work \cite{shao2010,camley2013}. The latter force results in 
cell shapes with roughly constant area. More details of these forces, 
along with details of the simulation techniques for Eqns. (\ref{phase_field}\&\ref{Stokes}) are given in Supplemental Material. 
As a consistency check, we have simulated cells without any propulsive force 
and have verified that the resulting static shapes  agree well with shapes
obtained using  standard energy minimization simulations \cite{brakke1992-s} (Fig. S3).

Polarization in our model is introduced through the 
	polarization indicator $\rho_a$ which is steering the actin polymerization. 
	For simplicity, we have chosen $\rho_a=1$ at the front half and $\rho_a=0$ at the rear half of the cell. This corresponds to different actin promoter (e.g., Rac or Cdc42) distributions at the front and back induced by internal or external signaling.
We assume the density of newly-made actin filaments is uniform at the cell membrane so we do not track the evolution of the actin density. In more complicated models\cite{shao2012,camley2013} the actin can diffuse or be advected but we do not include them here to keep the model simple.
	Following earlier work \cite{kruse2006,nagel2014,tjhung2015minimal}, we assume that   protrusions that are generated by actin polymerization only occur at the front of the cell and close to the substrate. Our formulation of the active 
	stress $\sigma^a$ incorporates these assumptions. 
	Specifically, we introduce a field $\psi(\bm{r})$  
with width $\lambda$ and
located a distance $\epsilon$ away from the substrate (Fig. S1). 
By making the active stress proportional to $G(\psi) \phi\rho_a(\bm{r})$, 
we restrict possible protrusions to a narrow  
 band  parallel to the substrate and in the cell front. This band is  schematically shown in yellow in Fig. \ref{fig1}.
In
addition, we  localize the stress to the interface  by multiplying the 
expression of $\sigma^a$ by the factor  $|\nabla\phi|^2$. 
This is schematically shown as blue dots in Figs. \ref{fig2} and 
\ref{act_stress}. 
The expression for the  stress is then given by:
\begin{equation} \label{act_stress}
\sigma^a=-\eta_aG(\psi)\phi\rho_a(\bm{r})\epsilon|\nabla\phi|^2\hat{\bm{n}}\hat{\bm{n}}.
\end{equation}
Here,   $\eta_a$  is the protrusion coefficient, and $\hat{\bm{n}}=\nabla\phi/|\nabla\phi|$ is the normal to the cell boundary.
Note that our model does not include any possible feedback between 
substrate and stress generation.

Our simulations are carried out as described previously \cite{camley2013} and 
 	further detailed in the Supplemental Material 
 	where we also list the full set of equations. 
	As initial conditions, our simulations start with polarized cells in which the distribution of $\rho_a$ is asymmetric. The cell's  speed is tracked by $\bm{v}_c=d\bm{x}_c/dt$ with $\bm{x}_c$ the cell mass center $\bm{x}_c=\int \bm{x}\phi d^2\bm{r}/\int\phi d^2\bm{r}$ 
and simulations are continued until a steady state has been achieved. 
Parameters values used in the simulations are given in Table S1.

\subsection{Simulation results and analysis}
\begin{figure}
\centering
\includegraphics[width=0.45\textwidth]{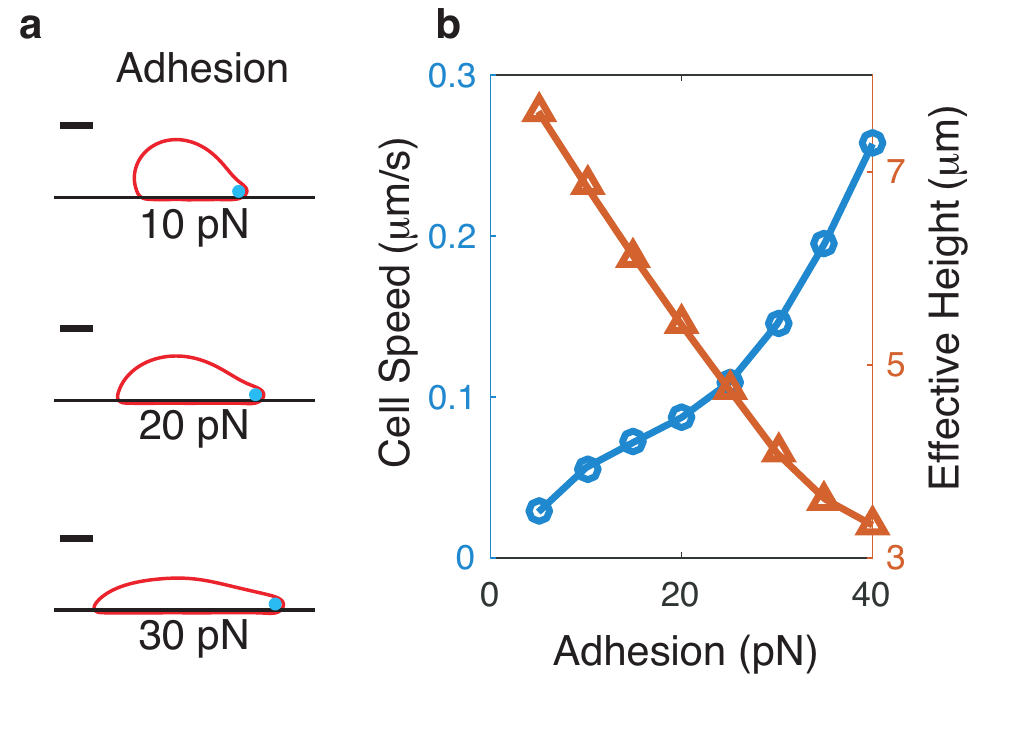}
\caption{\textbf{a}, Cell shapes for different values of substrate adhesion strength. The blue dots here, and elsewhere, schematically indicate the location of active stress. Scale bar $5\mu m$.  \textbf{b}, Cell speed (blue circles) and effective height  of a cell (red triangles) as a function of the adhesion strength. }
\label{fig2}
\end{figure}
We first investigate   how cells move on a single substrate with different adhesion energies. 
For this, we solve the 
phase field equations for different values of the adhesion 
parameter $A$.  Examples of resulting cell shapes are shown in 
Fig. \ref{fig2} while an example of the
actin fluid velocity field is shown in Fig. S4.
We find that with increasing  adhesion strength, cells spread more and thus become thinner, similar to the spreading of a droplet on surfaces with increasing wettability  (Fig. \ref{fig2}a). Our simulations reveal that the cell speed (i.e., the velocity parallel to the substrate) keeps increasing as the adhesion increases, without any indication of saturation (Fig. \ref{fig2}b). 
This is perhaps surprising, as our
physical intuition suggests that adhesion and friction go hand in
hand,  with larger adhesion corresponding to higher friction.
In our simulations, however, adhesion and 
friction are independent and can be separately adjusted.

To provide insights into the relation between adhesion, cell shape and speed, we consider a simplified version of Eq. (\ref{Stokes}), similar to  the 1D model
examined in Ref. \cite{carlsson2011}.
 Since only asymmetric stress will contribute to the cell's speed \cite{tanimoto2014}, we only need to take into account  the viscosity, friction and active stress in the equation:
 \begin{equation} \label{simple_Stokes}
 \nu \nabla\cdot\sigma^{vis}-\xi\bm{u}+\nabla\cdot\sigma^{a}=0,
 \end{equation}
 where  $\sigma^{vis}=\nabla\bm{u}+\nabla\bm{u}^T$, 
 $\xi$ is a friction coefficient taken to be spatially homogeneous, and $\sigma^a$ is the active stress which is 0 outside the cell.
Boundary conditions include  a steady cell shape $\hat{\bm{n}}\cdot\bm{v}_c=\hat{\bm{n}}\cdot\bm{u}$, zero net traction force $\int\xi\bm{u}d^2\bm{r}=0$, and zero parallel stress $\sigma^{vis}\cdot\hat{\bm{t}}=0$, where $\hat{\bm{n}},\hat{\bm{t}}$ are the normal and tangential unit vector, respectively.

It is in general not possible to solve Eq. \ref{simple_Stokes} in a arbitrary geometry. However, for the special case of a fixed-shape rectangular cell with length $L$ and height $H$ occupying  $x\in[-L/2,L/2],y\in[0, H]$  
we can solve for the cell speed $v_c$ (see the Supplemental Material).  
By averaging the stress over the vertical direction and following Carlsson's one-dimensional solution  \cite{carlsson2011}, we find:
\begin{equation} \label{ux}
v_c=-\frac{1}{4\nu H}\int_{-L/2}^{L/2}\frac{\tilde{\sigma}^{a}_{xx}\sinh(\kappa x)}{\sinh(\kappa L/2)}dx,
\end{equation}
where $\kappa=\sqrt{\xi/(2\nu)}$ 
 determines the spatial scale of the decay of a point 
stress source \cite{carlsson2011} and where 
$\tilde{\sigma}^{a}_{xx}=\int_0^H\sigma^{a}_{xx}dy$ (see also the Supplemental Material).
From this solution it is clear that asymmetric active stress distribution will lead to cell motion. When $\kappa L/2\ll1$, corresponding to a 
highly viscous cytoskeleton \cite{Bauetal98}, the speed is proportional to the normalized active stress dipole $1/(LH)\times\int x\tilde{\sigma}_{xx}dx$. 
In the phase field model, 
the active stress in Eq. \ref{act_stress} is a negative bell shape function located at the front tip of the cell. This active stress can be approximated by  $\tilde{\sigma}_{xx}=-\lambda\beta\delta[x-(L/2)_-]$ where $\beta$ is the active stress strength and where the stress is assumed to be 
located just inside the cell (see the Supplemental Material and Ref.  \cite{carlsson2011}). Substituting this  into Eq. \ref{ux}, we find 
\begin{equation}\label{vc_scale}
v_c=\frac{\lambda\beta}{4\nu H},
\end{equation}
which shows that the cell speed scales inversely with the height of the cell, and that  this scaling is independent of the cell length. Of course, a real cell will not be rectangular, and in the Supplemental Material we show that the cell speed scales with the average height for a more complex-shaped cell
(Fig. S5). 
This suggests that the cell speed can be parameterized using
 an  effective height $H_\text{eff}$, which  
can be computed by averaging the height over the cell length: $H_\text{eff}=(1/L)\int H(x)dx$. In Fig. \ref{fig2}b we see that $H_\text{eff}$ is monotonically decreasing when adhesion increases. The inverse relation between cell speed and effective height qualitatively agrees  with the above analysis. 

\begin{figure*} \centering
\includegraphics[width=.8\textwidth]{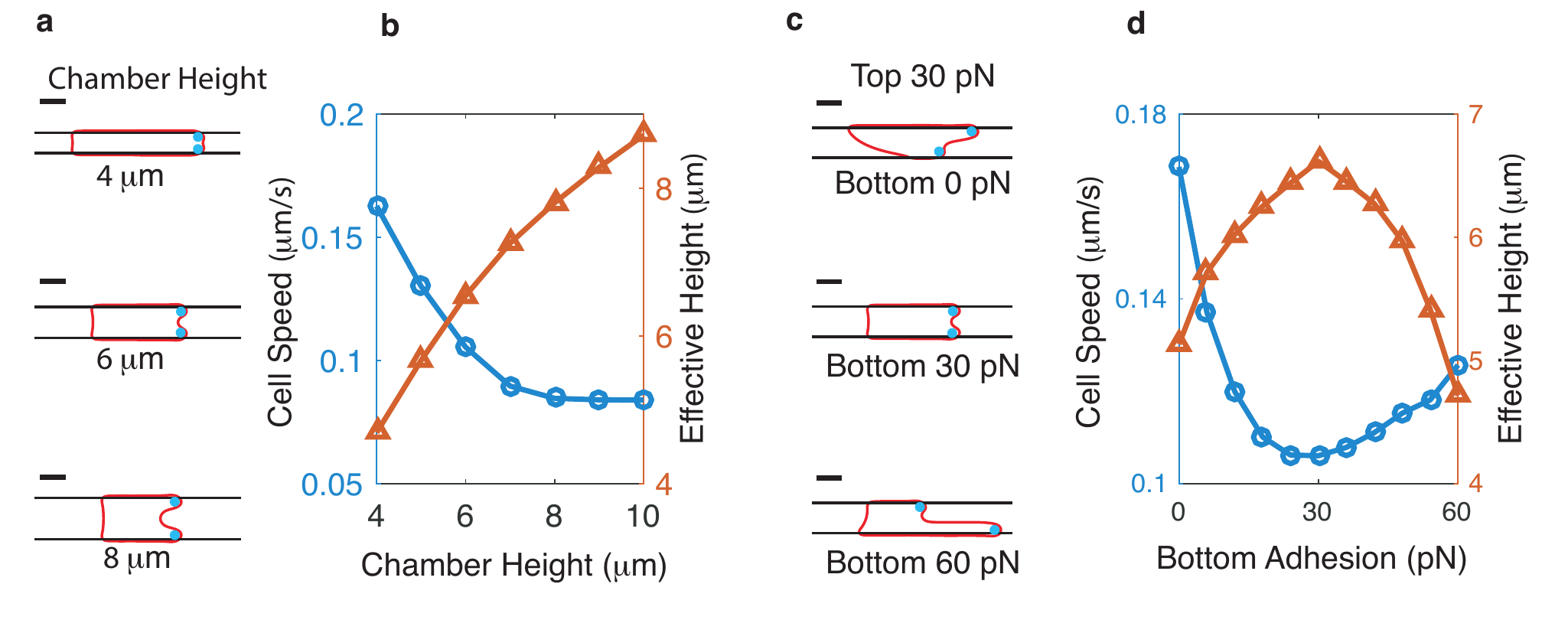}
\caption{\textbf{a}. Cell shapes for different chamber heights. The adhesion strength of the top and bottom substrate is fixed at 30 pN. Scale bar: $5\mu m$. \textbf{b}. Corresponding cell speed (blue circles) and effective height (red triangles) as  a function of chamber height.  \textbf{c}. Cell shapes in a chamber with adhesive  top and bottom substrates, with the top substrate  adhesion fixed to 30 pN.  Scale bar: $5\mu m$. \textbf{d}. Cell speed (blue circles) and effective height (red triangles) as a function of adhesion strength of the 
	bottom substrate (chamber height=6$\mu$m). }
\label{confined}
\end{figure*}

Interestingly, the above found relation between cell speed and 
cell height does not  depend on the way the cell's effective height 
is altered. To verify this, we also simulated cells in 
confined geometries in which they are ``squeezed" between two 
substrates, as shown in 
Fig. \ref{confined}a (an example of a cell with 
the actin fluid velocity field can be found in Fig. S4).
Consistent with our analytical results, we find that 
as the chamber height is reduced,  the cell's speed increases while the cell's effective height
decreases (Fig. \ref{confined}b). 
Furthermore, changing the adhesive strength on the   top and bottom substrate while  keeping the distance between them  fixed will also affect the 
cell shape and its effective height (Fig. \ref{confined}c and Fig. S4).
Our simulations show that a difference in the top and bottom adhesion leads to an asymmetric cross-section and that the cell's effective height reaches a maximum for equal top and bottom adhesion (Fig. \ref{confined}d). Consistent with Eq. \ref{vc_scale}, our simulations show that the cell speed reach a minimum for substrates with equal adhesive strengths (Fig. \ref{confined}d).

Our results can be explained by realizing that 
 cells contain a cytoskeleton network  that can be described as a compressible viscous actin fluid.
 This actin fluid contains ``active" regions which are confined to a layer with fixed width of $\lambda$, and ``passive" regions that are outside these active regions. Active stress is only generated 
 within this active region. Large viscosity will make the cell speed independent of cell length (see Eq. \ref{ux} and Ref. \cite{carlsson2011}). However, this viscosity also leads to dissipation due to internal shear stress: passive regions are coupled to the active regions through vertical shear interactions, resulting in dissipation. 
  This  dissipation increases with increasing cell height, as can also be seen in the velocity profile shown in Fig. S6, and  thinner cells will move faster.
 We have tested this explanation by carrying out additional simulations. 
 In one set, we simulated cells moving in chambers of varying height 
 while keeping the ratio of the 
size of the active stress layer
$\lambda$ and  cell height constant. Consistent with our theoretical
 predictions, the speed of these cells is independent of the chamber 
 height (Fig. S7). In addition, we have simulated cells in which the active stress region spans the entire front. Again in line with our theoretical 
 insights, the cell speed was found to be largely independent of the 
 chamber height (Fig. S8).
 
 \begin{figure*} \centering
\includegraphics[width=.8\textwidth]{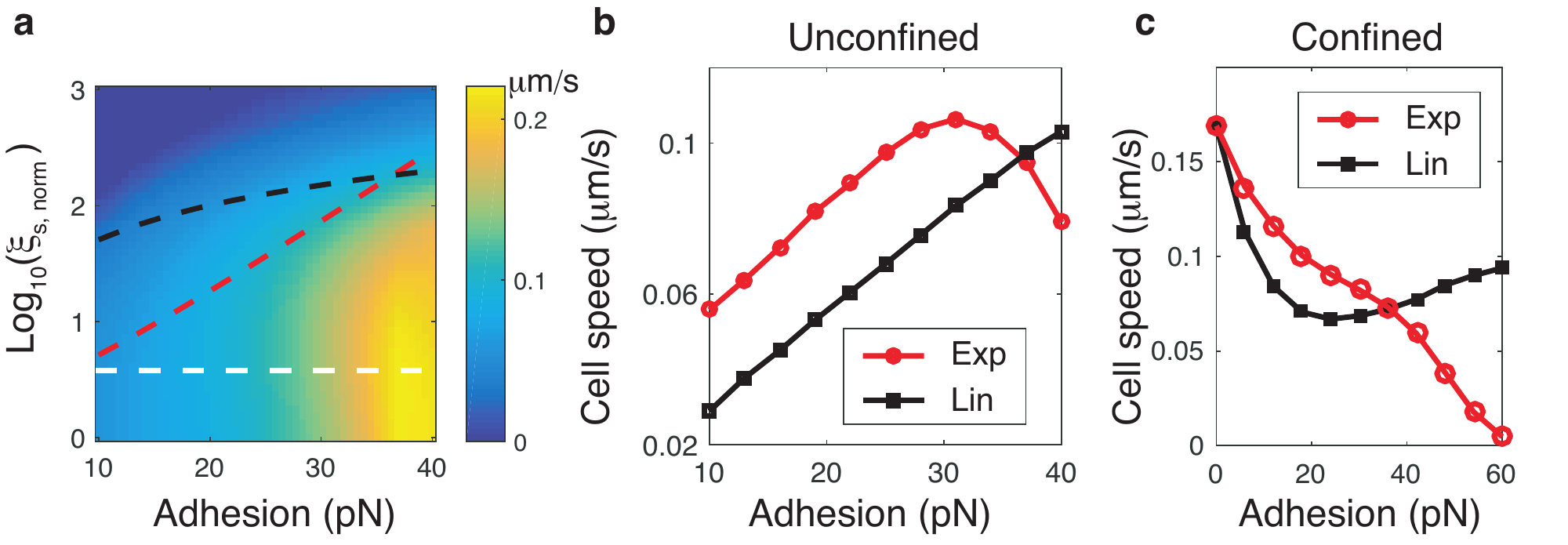}
\caption{\textbf{a}, Cell crawling speed dependence on adhesion strength and 
	friction coefficient (normalized by $\xi_s=$1 Pa s/$\mu$m). 
	Cell speed is visualized using the colormap. The dashed lines correspond to different dependencies of the friction on  adhesion
	(white: constant friction, black: linear dependence, 
	red: exponential dependence). \textbf{b}, Cell speed for unconfined cells as a function of adhesion for linear (black line) and exponential (red line) dependence on adhesion (for parameters see main text).  \textbf{c},  Speed of confined cells as function of bottom substrate adhesion strength for  linear (black line) and exponential (red line) dependence of friction on adhesion  (top substrate  adhesion=30 pN, chamber height=6$\mu$m). }
\label{fig4}
\end{figure*}

	In our  simulations, we have kept the friction coefficient constant and 
	have thus ignored any potential link between adhesion and friction.
	This is likely appropriate for {\it Dictyostelium} cells but may not be 
	 valid for mammalian cells that have integrin mediated focal adhesions.  
The exact dependence  of friction on  adhesion is  complicated and 
poorly understood \cite{srinivasan2009,walcott2010}. Our model, however,
 can easily be extended to explore    the entire phase space of friction and adhesion. To illustrate this,  we compute the speed of a cell crawling on 
 a single substrate by sampling a broad range of adhesion strengths ($A=10$ pN to $A=40$ pN) and friction coefficients 
 ($\xi_s=1$ Pa s/$\mu$m to $\xi_s=10^3$ Pa s/$\mu$m) while keeping
 all other parameters fixed. The  resulting cell speeds are shown in Fig.\ref{fig4}a
 using a color map. As expected, cells  stall when adhesion is low and friction is high (dark blue region) while the highest cell speed  occurs for large adhesion and a relatively broad range of low friction
 values (yellow region).

 Different dependencies between friction and adhesion correspond 
 to different trajectories through the two-dimensional phase space
 of Fig. \ref{fig4}a. 
 The results we have presented so far correspond to 
 traversing the phase space along the white dashed line in Fig. \ref{fig4}a.
 The black dashed line in this figure, on the other hand, represents a linear dependence between friction and adhesion ($\xi_s=
 \xi_b+\xi_l A/A_l$ with $\xi_b=$1 Pa s/$\mu$m, $\xi_l=$5 Pa s/$\mu$m, and $A_l=$1 pN) while the red dashed line represents an exponential 
 dependence ($\xi_s=\xi_b +
 \xi_e\exp{(A/A_e)}$ with $\xi_b=$1 Pa s/$\mu$m, $\xi_e=$1 Pa s/$\mu$m,  and $A_e=$7 pN).   
 For these two adhesion-friction dependencies, we have computed the cell 
 speed for unconfined (Fig.\ref{fig4}b) and confined cells (Fig.\ref{fig4}c). 
 For friction that depends linearly on adhesion, 
  the speed of unconfined cells  continues to  increase as adhesion increases (black line in Fig.\ref{fig4}b). 
  This is very similar to the results we obtained for constant friction (cf. Fig. \ref{fig2}b). 
  For exponential friction, the speed 
  of unconfined cells initially increases for increasing adhesion. 
  As adhesion increases, however, friction becomes more and more dominant, 
  and  cell's speed reaches a maximum, followed by a decrease (red line
  in Fig.\ref{fig4}b). This bi-phasic dependence of adhesion is consistent with
  a variety of  experiments\cite{huttenlocher1996,palecek1997integrin,gardel2008,barnhart2011}.
  For confined cells and a linear friction-adhesion relationship, the dependence of the cell speed on the adhesive strength of the 
  bottom substrate is shown in Fig. \ref{fig4}.c (black line). Again, 
  the results are very similar to our previously studied, constant 
  friction case (cf. Fig.\ref{confined}d): cell speed reaches a minimum when the top and bottom adhesion strength are equal. Not surprisingly, 
  the dependence of cell speed on bottom adhesion is different for 
  the exponential relationship. Here, friction becomes dominant when adhesion increases, resulting in a  cell speed that continuously  decreases. 
  These results show that 
  friction plays a relatively small 
  role  in determining cell speed unless friction $\xi_s$
  increases over orders of magnitude when adhesion $A$ changes by small
  amounts.

\begin{figure*}\centering
\includegraphics[width=.8\textwidth]{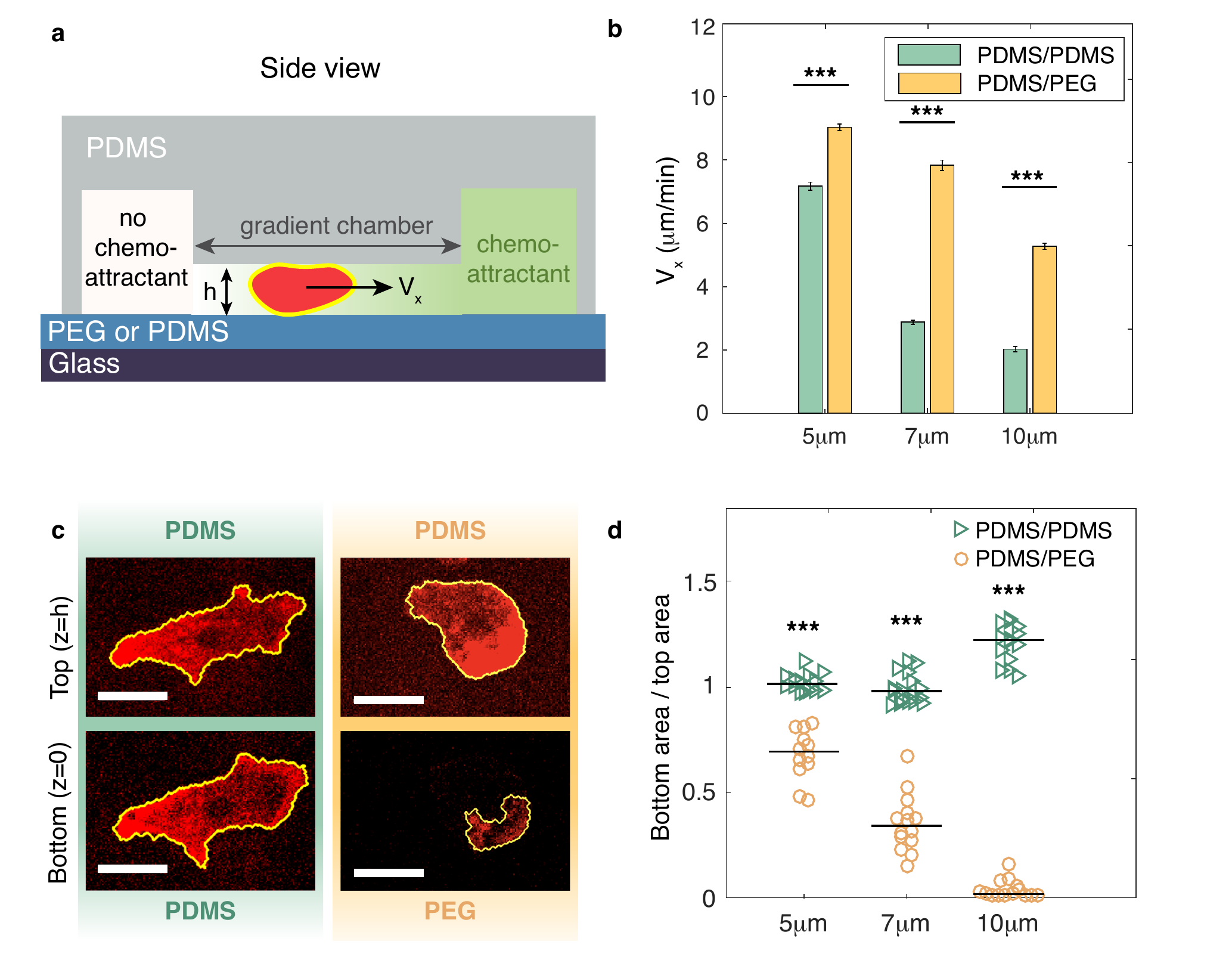}
\caption{Experimental tests of the numerical and theoretical predictions. \textbf{a}. Schematic side view of the microfluidics chamber.  Cells are placed in a confined chamber with variable height. The top substrate is composed of PDMS while the bottom substrate is either composed of 
		PDMS or coated with less adhesive PEG. Cells are guided by a chemoattractant (cAMP) gradient of strength 0.45 nM/$\mu$m in the chamber. \textbf{b}. Cell speed in the gradient direction for varying chamber height, indicated along the x-axis, and top/bottom substrate composition,  indicated by the label. For both PDMS/PDMS and PDMS/PEG substrate compositions, the cell speed decreases as  chamber height is 
	increased. Furthermore, for fixed chamber height, cells move faster when the bottom substrate is less adhesive (i.e., PDMS/PEG). 
	  P-value $<10^{-5}$ with unpaired t-test. Error bars represent the standard error of the mean.  \textbf{c}. Scan of the cell area profile at the top and bottom of the chamber using the fluorescent membrane marker Car1-RFP ($h=5 \mu$m). Cells with PEG-coated bottom substrates (PDMS/PEG) show asymmetric shapes whereas cells with PDMS bottom substrates (PDMS/PDMS) show symmetric shapes. Scale bar, 10$\mu$m.  \textbf{d}. Ratio of  top    and bottom contact area under different conditions (label indicates bottom substrate composition).  P-value $<10^{-5}$ with Wilcoxon rank-sum test. }
\label{exp}
\end{figure*}

\subsection{Experimental results}
To test the above predictions, we performed motility experiments of \textit{Dictyostelium discoideum} cells. 
Importantly, these cells, unlike mammalian cells, do not make integrin mediated focal adhesions and their substrate adhesion is likely to be mediated by direct physiochemical factors such as van der Waals attraction 
	\cite{loomis2012innate}.
Experiments are carried out in microfluidic devices, as shown in Fig. \ref{exp}a and modified from earlier work \cite{Skoetal14} (see also 
Supplemental Material and Fig. S9). 
Cells are moving in chambers with  height  $h$ and 
with substrates that have variable adhesive properties. A constant cAMP gradient is established by  diffusion so that  cells preferably move in one direction (denoted as the $x$ direction). 
Note that the constant signal polarizing the cell
in one direction is consistent with our model of a
constantly-polarized cell.

\textit{Dictyostelium} cells move by extending
 	actin filled protrusions called pseudopods which can extend over a significant distance from the substrate. 
 	As a consequence, our confined cells occlude the entire 
 	space between two substrates. This  was verified explicitly by 
 	labeling the cell with a fluorescent membrane marker and 
 	creating confocal z-stacks (Fig. S10). The results also demonstrate that in the case of symmetric adhesion the outline of the cell does not change appreciably as one moves from one to the other substrate. Furthermore, using LimE as a fluorescent marker, we have  verified that the level of actin polymerization is largest near the substrates (Fig. S10 C). This observation is in agreement with earlier experiments of   {\it Dictyostelium} cells migrating in a narrow channel \cite{nagel2014} which revealed significantly larger levels of  LimE fluoresence near the channel walls. 
 	%In contrast to the findings in\cite{liu2015}, where the cells form elongated pseudopodia under confinement, we find the cells in our experiments maintain a smooth outline (Fig. S10).

	The  top substrate of the chamber consists of Polydimethylsiloxane (PDMS) and the  bottom substrate is either made of PDMS or is coated with a thin layer of  Polyethylene glycol  (PEG)  gel.
 	Cells moving 
on these PEG-coated substrates have vastly reduced adhesion, as 
reported in earlier studies \cite{Tzvetal09}. 
We measure the average speed of the cell $V_x$ in the direction of the 
chemoattractant gradient, both as a function of the 
height of the chamber and for different substrate compositions  (Fig. \ref{exp}b). Furthermore,
to quantify the effect of the adhesive properties of the substrates on 
migrating cells, we measure the contact area of the cell on both  top and bottom substrates of the chamber using confocal microscopy (Fig. \ref{exp}c and d).
More adhesive substrates will result in more cell spreading and thus larger 
contact areas.

 Our theoretical predictions for cells in confinement
 are that decreased height increases speed, and that cells in
 asymmetric adhesion are faster than cells in symmetric adhesion. Both
 of these qualitative predictions are observed in our experiments.
 First,
our experiments show that cell speed is significantly affected by the 
 height of the chamber (Fig. \ref{exp}b). Cells in chambers of height $h=10\mu$m move markedly slower than cells in chambers with 
 $h=7\mu$m which, in turn, have smaller speed than cells in chambers with 
 $h=5\mu$m.
 The trend of slower motion in deeper chambers holds for both PDMS and PEG coated bottom substrates. 
 Furthermore, we have verified that these results do not depend on the 
 steepness of the gradient (Fig. S11).
 These observations are  fully consistent with our numerical and theoretical predictions (Fig. \ref{confined}).

  In addition, our experiments show that cells moving in a chamber
  with unequal top and bottom adhesion are markedly asymmetric
  (Fig. \ref{exp}c), consistent with past results that showed that
  {\it Dictyostelium} cells only weakly adhere to PEG. 
 Specifically, the contact area of cells on PEG 
coated substrates is significantly smaller than the contact area 
on PDMS substrates and the resulting asymmetry can be quantified by the 
ratio of bottom and top contact area. 
Cells with PDMS  on top and bottom and for $h=$5$\mu$m and 
$h=$7$\mu$m have ratios close to 1 indicating 
that the shape is symmetric. In contrast, cells moving in chambers with 
these values of $h$ that have 
a  PEG bottom have ratios that are much smaller than 1,
indicating a more asymmetric cell shape. 
For the largest value of $h$ ($h=$10$\mu$m) cell preferentially attach to the top PDMS substrate, resulting in negligible contact area 
at the bottom PEG substrate and ratios close to 0.
For this chamber height, the  ratio for PDMS substrates is
 larger than one since cells are loaded on the bottom substrate and cannot fully attach to the top substrate.

Importantly, quantifying the cell speed for the different chambers reveals that
 cells in the symmetric PDMS/PDMS condition move slower than
 cells in the asymmetric PDMS/PEG condition
 (Fig. \ref{exp}b).
Again, these 
experimental results are fully consistent with our theoretical and numerical predictions and show that  cell shape, and more specifically its effective 
height, can significantly   affect  motility speed (Fig. \ref{confined}).

\section{Discussion and conclusion}
In this study, we examined how cell shape can affect 
	cell speed using simulations, analytics, and experiments.
We should stress that our experiments can only be compared to the 
simulations on a qualitative level. 
Values for the model parameters 
 are not precisely known, and our model cell 
is not fully three-dimensional. Nevertheless, separating the 
frictional and adhesive force in the model provides 
clear insights into the role of adhesion and cell shape in determining 
cell speed. 
This separation also makes it  challenging to compare our results to 
previous studies that investigated the effects of cell-substrate interactions
on cell speed. 
For example, a recent study using fish keratocyte cells \cite{barnhart2011}  found that
 cell spreading  increases with adhesion strength (measured by the 
 concentration of adhesive molecules).
 These experiments also 
 revealed a biphasic speed dependence on adhesion such that cell speed increases between low and
 intermediate adhesion strengths and decreases between intermediate
 and high adhesion strengths. These results are similar to
 earlier experimental studies, and have previously been interpreted in
 terms of minimal models without cell shape
  \cite{Dimetal91,palecek1997integrin,liu2015}. 
Our results suggest that the increase of cell speed with increased adhesion found in these experiments might 
be attributed to cell spreading and a lower effective height.
The observed decrease in cell speed following a further
increase in adhesion can then be explained by a larger relative 
role of frictional forces. 
Likewise, our experimental results suggest that 
our experiments operate in a regime where substrate friction is less important than the internal viscosity and hence the major effect of the substrate modification is the change in adhesion. 

Our numerical and experimental results indicate that changing cell morphology through confinement can also significantly alter the migration speed, with decreasing chamber heights resulting in increased cell speeds. Comparison with other cell types is challenging as cells might change their behavior following confinement. 
	 A recent study using normal human dermal fibroblast cells, for example,  found that slow mesenchymal cells can spontaneously switch to a fast amoeboid migration phenotype under confinement \cite{liu2015}. This phenotypic  transition makes it difficult to directly compare those observations with our results and further investigation is needed to determine how different cell types behave in confinement.

We should point out that the simple scaling of cell speed dependence on cell height (Eq. \ref{vc_scale}) is based on the assumption of localized active stress (the numerator) and uniform cytoskeleton viscosity (the denominator) in the entire cross section. 
	As shown in our experimental work and in previous studies \cite{nagel2014}, 
	F-actin is localized close to the substrate, in support of the 
	first assumption. It is currently unclear whether the second approximation is valid for {\it Dictyostelium} or other cells. Presumably, in cells with a clear segregation of actin cortex and cytoplasm,   a large viscosity contrast could be present. Nevertheless, our arguments  might still hold, as long as  passive regions are coupled to the active regions  through shear interactions (one example is the model in Ref.\cite{kruse2006}). In this case,  passive regions will still slow down the cell, but the relation between cell speed and shape will be more complicated and will have  contributions from regions with different viscosity. 
	To address this  more general case, a full three-dimension model with viscosity contrast between the cortex and cytoplasm is necessary and will be part of future extensions.

In summary, we show how adhesion forces result in cell spreading and that 
the accompanying shape changes can result in larger velocities. 
Key in this result is the existence of a narrow band of 
active stress that has a smaller spatial extent than the height of the cell. 
As a result, the dissipation due to the shear stress between this active band and the remainder of the cell increases as  
the effective height of the cell increases. 
 In our model, we have assumed a cell motility model corresponding to 
 stable flat protrusions. The conclusion that cell speed scales inversely with the effective height is also valid for other cell motility models as long 
 as the active propulsion region has limited spatial extent. For example, 
 replacing the constant active stress by an oscillating stress, 
 similar to  protrusion-retraction cycles seen in amoeboid cells, does not 
 change the qualitative results (Fig. S12). Further extensions of our 
 model could include focal adhesive complexes (to model
a broader range of eukaryotic cell types) and different types of actin structures in different parts of the cell.  These extensions can then be used to further 
 determine the role of adhesion in cell motility.

% 

%%%END OF MAIN TEXT%%%



\pagebreak
\widetext
\begin{center}
\textbf{\large Supplemental  Material for ``Cell Motility Dependence on Adhesive Wetting"}
\end{center}
%%%%%%%%%% Merge with supplemental materials %%%%%%%%%%
%%%%%%%%%% Prefix a "S" to all equations, figures, tables and reset the counter %%%%%%%%%%
\setcounter{equation}{0}
\setcounter{figure}{0}
\setcounter{table}{0}
\setcounter{page}{1}
\makeatletter
\renewcommand{\theequation}{S\arabic{equation}}
\renewcommand{\thefigure}{S\arabic{figure}}
\renewcommand{\bibnumfmt}[1]{[S#1]}
\renewcommand{\citenumfont}[1]{S#1}

\section{Phase field model of cell motility}
The equations for the phase-field cross section model are:
\begin{align}
&\frac{\partial\phi(\bm{r},t)}{\partial t}=-\bm{u}\cdot\nabla\phi(\bm{r},t)+\Gamma(\epsilon\nabla^2\phi-G^{\prime}/\epsilon+c\epsilon|\nabla\phi|) \\
&\nabla\cdot[\nu \phi(\nabla\bm{u}+\nabla\bm{u}^T)]+\bm{F}_{sub}+\bm{F}_{mem}+\bm{F}_{area}+\nabla\cdot\sigma^a=0.
\end{align}
Here, $\phi$ describes the field of the cell. The  double-well potential is defined as  $G=18\phi^2(1-\phi)^2$ and the 
curvature is computed as $c=-\nabla\cdot(\nabla\phi/|\nabla\phi|)$ while 
$\Gamma$ is a relaxation coefficient. The force terms are explicitly explained below.

\begin{figure} \centering
\includegraphics[width=.45\textwidth]{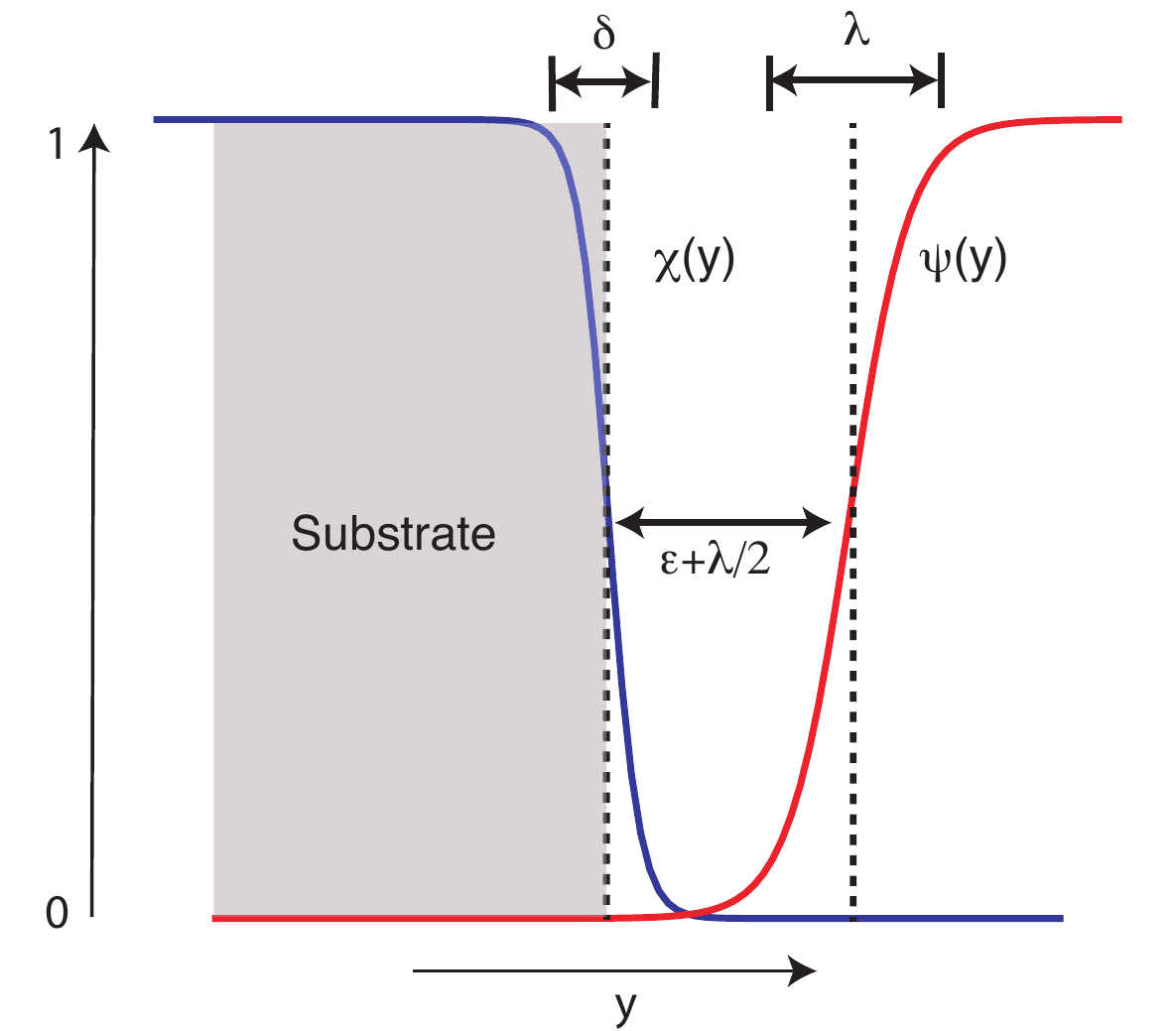}
\caption{Illustration of the substrate field $\chi$, with width $\delta$, together with the 
		protrusion band $\psi$, with width $\lambda$ and located a distance 
		of $\epsilon$ away from the substrate.}
\label{schem}
\end{figure}

The substrate force contains the cell-substrate adhesion and friction: $\bm{F}_{sub}=\bm{F}_{adh}+\bm{F}_{fric}$, where 
\[
\bm{F}_{fric}=-\xi_s\chi\bm{u} -\xi_d\bm{u},\;\bm{F}_{adh}=\frac{\delta H(\phi,\chi)}{\delta \phi}\nabla\phi.
\]
Here, $\bm{u}$ is the velocity field  of the actin fluid and
$\xi_s,\xi_d$ are the cell-substrate friction coefficient and damping coefficient, respectively. $\chi$ is the field describing the substrate, and $H(\phi,\chi)$ is the interaction potential between the cell and substrate. The The cell moves either on top of a plain substrate or between a top and bottom substrate. The location of these substrates is given by a  field $\chi(y)$ 
with a boundary width of $\delta$ (Fig. \ref{schem}).
Here, $\chi=1$ indicates the substrate into which  the cell cannot penetrate, and $\chi=0$ indicates the region accessible to the cell. In our simulations, the substrate is parallel to the x direction and, for the case of a 
single substrate located at $y=y_B$, 
$\chi$ is written as 
\[
\chi(y)=\frac{1}{2}-\frac{1}{2}\tanh\{3(y-y_B)/\delta\},
\]
For a chamber with a  parallel top substrate located at 
$y=y_T$ this becomes  
\[
\chi(y)=\frac{1}{2}+\frac{1}{2}\tanh\{3[|y-(y_T+y_B)/2|-(y_T-y_B)/2]/\delta\}.
\]
Given $\phi$ and $\chi$, the interaction potential is:
\[
H(\phi,\chi)=\int d\bm{r}^2\phi^2(\phi-2)^2 W(\chi),
\]
where $W(\chi)$ contains an attractive term, corresponding to adhesion, and a repulsive term, corresponding to the non-penetrability of the substrate.
For the bottom substrate, we use
\begin{equation}
W(\chi)=-2A\frac{G(\chi)}{\delta}+\frac{g}{2}\chi(y+\epsilon),
\end{equation}
while the potential for the top substrate has an identical form with 
$\epsilon$ replaced by $-\epsilon$. Here, $A$ is the adhesion energy per unit length, $g$ is a parameter that 
measures  the penalty  of overlap between cell and substrate \cite{camley2014-s}, 
and $G$ is a double-well potential $G(\chi)=18\chi^2(1-\chi)^2$.
The energy function 
\[
H(\phi,\chi)=\int \phi^2(\phi-2)^2W(\chi)d^2\bm{r}
\]
corresponds, in the sharp interface limit, to 
an adhesive energy equal to $-A l$ where $l$ is the length of 
the cell in contact with the substrate.
Note that the inclusion of the  $\phi^2(\phi-2)^2$ results in  a 
force that only vanishes outside the membrane \cite{zhao2017new-s}. 
In our simulations we take $\delta=\epsilon/2$. For this choice of 
$\delta$ we  simulated cells without any propulsive force. The resulting 
static shapes can be directly compared to  standard energy minimization simulations. Fig. \ref{wetting}
shows that the phase field shapes  
agree well with shapes
obtained using  Surface Evolver, 
a simulation tool that evolves surfaces toward minimal energy by a gradient descent method \cite{brakke1992-s}.

The contribution from both the tension and bending of the membrane is captured by $\bm{F}_{mem}$. 
 In our simulation we ignore the bending term since it contributes little to the shape of cell. The tension energy is given by \cite{shao2010-s,camley2013-s}:
\[
H_{ten}=\int\frac{\gamma}{2}[\epsilon|\nabla\phi|^2+\frac{G(\phi)}{\epsilon}]d^2\bm{r},
\]
resulting in $\bm{F}_{mem}=\frac{\delta H_{ten}}{\delta \phi}\nabla\phi$.
Area conservation is introduced via $\bm{F}_{area}=M_a(\int\phi d\bm{r}^2-A_0)\nabla\phi$ with $A_0$ the prescribed area size, and $M_a$ 
a parameter which controls the strength of the area constraint \cite{shao2010-s}.

The  active stress term in our model, 
$\sigma^a=-\eta_aG(\psi)\phi\rho_a \epsilon|\nabla\phi|^2 \hat{\bm{n}}\hat{\bm{n}}$, is similar to our earlier work \cite{shao2012-s} but only acts near the substrate. This is accomplished  through the addition of the 
term  
 $G(\psi)=18\psi^2(1-\psi)^2$, where $\psi$, for the bottom 
 substrate,  takes on the form 
\[
\psi_{B}(y)=\frac{1}{2}+\frac{1}{2}\tanh\{3[y_{B}+(\epsilon+\lambda/2)-y]/\lambda\}.
\]
A similar expression is used for the top substrate. 
The inclusion of $G(\psi)$  results in active stresses confined to  a band with width $\lambda$ and located a distance $\epsilon$ away from the substrate 
(Fig. \ref{schem}).
Note that vertical height of the active stress is controlled by 
$\lambda$ and that $\int G(\psi)dy=\lambda/2$. 

Three examples of the velocity fields obtained numerically are shown in Fig. \ref{flow}, corresponding to the cell motion on single substrate, confined in channels and confined in channels with asymmetric adhesion (Fig. 2 and Fig. 3 in main text). The retrograde flow patterns are similar to previous studies in\cite{shao2012-s}.

\section{Numerical Methods}

\vspace{5mm}
The equation for $\phi$  is stepped by uniform time step $\Delta t=2\times10^{-3} s$ in a forward Euler scheme so that 
  $\phi$ at time step $n+1$ is obtained from 
 $\phi$ at time step $n$: 
\[
\phi^{(n+1)}=\phi^{(n)}-\Delta t\bm{u}+\Delta t\Gamma[\epsilon\nabla^2\phi^{(n)}-G^{\prime}(\phi^{(n)})+\epsilon c^{(n)}|\nabla\phi^{(n)}|],
\] 
Here, $c^{(n)}=-\nabla\cdot(\nabla\phi^{(n)}/|\nabla\phi^{(n)}|)$ is computed using a finite difference method and all other differentiation operators  are  computed using a fast Fourier spectral method.  Simulations were carried 
out on a $256\times 256$ grid of size $50\mu m\times 50\mu m$. 
Model parameters, modified from \cite{shao2012-s,camley2013-s}, are listed
in Table S1.

The velocity field $\bm{u}$ is updated every time step by a semi-implicit Fourier spectral method after updating $\phi$ as detailed in 
\cite{camley2013-s}. The equation is iterated as:
\[
\xi_0\bm{u}_{k+1}-\nu\tilde{\phi}\nabla^2\bm{u}_{k+1}=\nabla\cdot[\nu\phi\nabla\bm{u}_{k}+\nu(\phi-\tilde{\phi})\nabla\bm{u}^T_{k}]-\xi_s\chi\bm{u}_k+\bm{F},
\]
where $\tilde{\phi}=2$, and $\bm{F}$ represents the terms in the Stokes equation that are independent of the iteration step $k$. The iteration will continue until
\[
\frac{\max|\bm{u}_{k+1}-\bm{u}_k|}{\max|\bm{u}_k|}<0.1,
\]
or until a maximal number of iterations (here chosen to be 20) is reached.

\section{Analytical Results}
As stated in the main text, we aim to analytically solve Eq. S1\&S2, where several simplifications have to be made. First, we are trying to find the steady-state solutions, so the cell shape will not change with time. Thus we drop  Eq. S1 and, instead, put boundary conditions for Eq. S2. In accordance with our simulations, we choose slip boundary conditions, similar to\cite{carlsson2011-s}. The boundary condition for the steady-state cell shape is
\[
\bm{u}\cdot\hat{\bm{n}}=\vec{v}_c\cdot\hat{\bm{n}},
\]
where $\vec{v}_c$ is the cell's mass of center velocity, which is our target to solve, and $\hat{\bm{n}}$ is the normal unit vector of the boundary. The cell's boundary is free so the parallel stress at the boundary is zero
\[
\hat{\bm{t}}\cdot\sigma^{vis}=0,
\]
where  $\hat{\bm{t}}$ is the tangential unit vector of the boundary. Notice that the active stress is always constrained inside the cell so it will not enter any boundary conditions. The total force of the cell exerted on substrate should be balanced which gives a zero net traction force condition
\[
\int\xi(\bm{r})\bm{u}d^2\bm{r}=0,
\]
where $\xi(\bm{r})$ is the friction coefficient at different locations. To get analytical expressions, we neglect the spatial heterogeneity in friction and simply take $\xi(\bm{r})=\xi$. This simplification does not change the central feature of our main result (the cell's speed is inversely related to the cell's height).

Second, we only take into account the viscosity, friction and active stress because they are directly related to the cell motion. The adhesion, area conservation and membrane forces only contribute to the cell's shape, which is implicitly included in the boundary conditions. Thus we get a simplified equation for Eq. S2:  
\begin{equation}
\nu\nabla\cdot\sigma^{vis}-\xi\bm{u}+\nabla\cdot\sigma^{a}=0.
\end{equation}
Integrating the above equation and using the zero traction force condition, we obtain $\oint (\nu\sigma^{vis}+\sigma^{a})\cdot\hat{\bm{n}}dl=0$. As the active stress $\sigma^a$ is constrained inside the cell, this will lead to a condition equivalent to the zero traction force condition
\[
\oint\hat{\bm{n}}\cdot\sigma^{vis}dl=0,
\]
which is the zero traction force condition we used below.

Notice that a fixed cell shape has to be  given in order to apply the boundary conditions. Since we only care about the cell's mass of center velocity $\vec{v}_c$, and not the full solution for $\bm{u}$, we 
will next show how $\vec{v}_c$ can be obtained without knowing $\bm{u}$.

\subsection{Analytical solution of the rectangular model cell}
Here we wish to solve the Eq. S4 for a rectangular fixed cell shape $x\in[-L/2,L/2],y\in[0, H]$ with an unknown cell speed $v_c$ (notice we put the x-direction as cell moving direction so $v_c$ is a scalar). The boundary conditions are  $\bm{u}_x(x=\pm L/2)=v_c,\;\bm{u}_y(y=0,H)=0,\; \rm{and} \int\bm{u}dxdy=0$. Integrating  the Stokes equation, we  get $\int d^2\bm{r}(\nu\nabla\cdot\sigma^{vis}+\nabla\cdot\sigma^{a})=\oint(\nu\sigma^{vis}\cdot\bm{n}+\sigma^{a}\cdot\bm{n})dl=\xi\int\bm{u}d^2\bm{r}=0$. Note that the active stress $\sigma^{a}$ should be constrained within the cell \cite{carlsson2011-s} resulting in the zero net traction force condition $\oint\sigma^{vis}\cdot\bm{n}dl=0$. This means $\int[\sigma_{xx}^{vis}(x=L/2)-\sigma_{xx}^{vis}(x=-L/2)]dy=\int[\partial_x\bm{u}_x|_{x=L/2}-\partial_x\bm{u}_x|_{x=-L/2}]dy=0$ and $\int[\sigma_{xy}^{vis}(y=H)-\sigma_{xy}^{vis}(y=0)]dx=0$ due to the rectangular shape.

The tangential vector $\hat{\bm{t}}$ can be determined by the normal vector $\hat{\bm{t}}_x=-\hat{\bm{n}}_y,\hat{\bm{t}}_y=\hat{\bm{n}}_x$. The zero-parallel stress condition $\hat{\bm{t}}\cdot\sigma^{vis}=0$ results in
\[
\hat{\bm{n}}_y\sigma^{vis}_{xx}-\hat{\bm{n}}_x\sigma^{vis}_{xy}=0,\;\;\hat{\bm{n}}_y\sigma^{vis}_{xy}-\hat{\bm{n}}_x\sigma^{vis}_{yy}=0.
\]
For rectangular boundaries, these  conditions lead to 
\begin{equation}
\sigma^{vis}_{xy}=0,
\end{equation}
at all boundaries.

Since the cell is moving along x-direction, only $\bm{u}_x$ is relevant and 
we can integrate the 2D Stokes equation in the y-direction. With the condition of $\sigma^{vis}_{xy}=0$, we obtain a 1D Stokes equation:
\begin{equation}
 \label{1D_Stokes}
2\nu\frac{\partial^2 \tilde{\bm{u}}_x}{\partial x^2}-\xi\tilde{\bm{u}}_x+\frac{\partial\tilde{\sigma} ^{a}_{xx}}{\partial x}=0,
\end{equation}
where $\tilde{\bm{u}}_x=\int_0^H\bm{u}_xdy$, and $\tilde{\sigma}^{a}_{xx}=\int_0^H\sigma^{a}_{xx}dy$. The corresponding boundary conditions are $\tilde{\bm{u}}_x(L/2)=\tilde{\bm{u}}_x(-L/2)=v_cH$ and $\partial_x\tilde{\bm{u}}_x|_{x=L/2}=\partial_x\tilde{\bm{u}}_x|_{x=-L/2}$. This is exactly the same problem as in reference \cite{carlsson2011-s}. Using standard Green's function methods, we obtain:
\[
\tilde{\bm{u}}_x(L/2)=-\frac{1}{4\nu}\int_{-L/2}^{L/2}\frac{\tilde{\sigma}^{a}_{xx}\sinh(\kappa x)}{\sinh(\kappa L/2)}dx,
\]
and, since $\bm{u}_x=v_c$ at boundaries $x=\pm L/2$, we obtain
\begin{equation}
v_c=\frac{\tilde{\bm{u}}_x(L/2)}{H},
\end{equation}
as reported in the main text (Eq. 5). 
 If the active stress is confined in a band with width $\lambda$, i.e., $\int_0^H\sigma^{a}_{xx}dy=\lambda f(x)$, the cell's speed $v_c$ will scale as:
\begin{equation}
v_c=\frac{\lambda v_0}{H},
\end{equation}
where $v_0$ is a  constant, corresponding to the boundary velocity determined by the 1D problem $2\nu\partial_x^2v-\xi v+f^{\prime}(x)=0$ with homogeneous boundary conditions.
Notice that this scaling does not depend on the vertical position of the active stress. Therefore, our model will give the same cell speed independent of the 
type of active stress (actin polymerization, myosin contraction), as long as the integrated active stress is the same.

\subsection{Effective height for non-rectangular cells} 

In the above section, the speed of a rectangular cell was determined exactly. Actual cells are, of course, not rectangular but obtaining a solution for 
cells with more complex shapes is challenging. Nevertheless, insight can 
be obtained by considering  a cell composed of two rectangles, one positioned at $[-L,0]\times[0,H_2]$ and one positioned at $(0,L]\times[0,H_1]$ ($H_1<H_2$ (see Fig. \ref{step}). We take the active stress to be located at the latter (right) rectangle. This problem has the same boundary conditions as above, with two additional continuity conditions:
\begin{equation}
\bm{u}_x(x=0^+)=\bm{u}_x(x=0^-),\;\;\;\partial_x \bm{u}_x|_{x=0^+}=\partial_x \bm{u}_x|_{x=0^-}.
\end{equation}
To simplify the problem, we introduce the new variables  $u_1=\int_0^{H_1} \bm{u}_xdy$ and $u_2=\int_0^{H_2}\bm{u}_xdy$.   Using the continuity condition we have:
\[
u_2(0^-)=\int_0^{H_2} \bm{u}_x(x=0^-)dy = \int_{H_1}^{H_2}\bm{u}_x(x=0^-)dy+\int_0^{H_1}\bm{u}_x(x=0^+)dy=(H_2-H_1)v_c+u_1(0^+).
\]
Together with $u_1(L)=H_1v_c, \, u_2(-L)=H_2v_c$ we get 
\begin{equation}
\left. u_2\right|_{-L}^0+\left. u_1\right|_0^L=0.
\end{equation}
The zero traction force will give 
\[
\int_0^{H_2}\partial_x\bm{u}_x|_{x=-L}dy=\int_{H_1}^{H_2}\partial_x\bm{u}_x|_{x=0}dy+\int_0^{H_1}\partial_x\bm{u}_x|_{x=L}dy.
\]
Combining with the stress continuity we obtain
\[
\partial_xu_2|_0=\int_0^{H_2}\partial_x \bm{u}_x|_{x=0}dy=(\int_0^{H_1}dy+\int_{H_1}^{H_2}dy)(\partial_x\bm{u}_x|_{x=0})=\partial_xu_1|_0+\partial_xu_2|_{-L}-\partial_xu_1|_L.
\]
such that
\begin{equation}
\left.\partial_xu_2\right|_{-L}^0+\left.\partial_xu_1\right|_0^L=0.
\end{equation}
Notice that Eq. S10 and Eq. S11 have clear physical meanings, namely  flow conservation and force balance, respectively. It is convenient to introduce the net flow $C$ and net force $F$ on each rectangle:
\[
u_2|_{-L}^0=C,\;\;u_1|_0^L=-C,\;\;\partial_x u_2|_{-L}^0=F,\;\;\partial_xu_1|_0^L=-F,
\]
and, using the zero-parallel stress condition, we obtain the 1D version of the problem for the right and left rectangle:
\[
2\nu\partial_x^2u_2-\xi u_2=0,\;\;\;2\nu\partial_x^2u_1-\xi u_1+\partial_x\sigma^a=0,
\]
with $\sigma^a=\int_0^{H_1}\sigma_{xx}dy$. $u_1$ can be solved by superposition of two parts: $\tilde{u}_1$ with homogeneous boundary conditions and active stress, and $\hat{u}_1$ with inhomogeneous boundary conditions but zero active stress.  After substituting $u_1=\tilde{u}_1+\hat{u}_1$, we obtain
\[
2\nu\partial_x^2\tilde{u}_1-\xi\tilde{u}_1+\partial_x\sigma^a=0,\;\tilde{u}_1|_0^L=0,\;\partial_x\tilde{u}_1|_0^L=0,
\]
and
\[
2\nu\partial_x^2\hat{u}_1-\xi\hat{u}_1=0,\;\hat{u}_1|_0^L=-C,\;\partial_x\hat{u}_1|_0^L=-F.
\]
We can then solve for  $\tilde{u}_1, \hat{u}_1$ and $u_2$ and obtain the boundary velocity:
\begin{equation}
H_1v_c=u_1(L)=v_a-\frac{C}{2}-\frac{\alpha F}{2\kappa},\;\; H_2v_c=u_2(-L)=-\frac{C}{2}+\frac{\alpha F}{2\kappa},
\end{equation}
and the boundary stresses:
\begin{equation}
\partial_xu_2|_{-L}=-\frac{F}{2}+\frac{\alpha\kappa C}{2},\partial_xu_2|_0=\frac{F}{2}+\frac{\alpha\kappa C}{2},\partial_xu_1|_0=\pi_a+\frac{F}{2}-\frac{\alpha\kappa C}{2},\partial_xu_1|_L=\pi_a-\frac{F}{2}-\frac{\alpha\kappa C}{2},
\end{equation}
where $\kappa=\sqrt{\xi/2\nu},\alpha=\coth(\kappa L/2)$. $v_a$ and $\pi_a$ are the boundary speed and boundary stress from the homogeneous equation of $\tilde{u}_1$, which are constants.

To calculate $v_c$, we have to determine $C$ and $F$. Eq. S12 gives one condition  $H_1/H_2=u_1(L)/u_2(L)$  and an additional condition from the stresses in Eq. S13 is needed. Unfortunately, there is no simple relation between the four equalities in Eq. S13 since the 
stress continuity equation cannot be defined at the boundary   at $x=0$ and between  $y=H_1$ and $y=H_2$. Instead, we assume that the ratio of the 
integrated stress at $x=0$ satisfies $(\partial_xu_1|_0)/(\partial_xu_2|_0)=\beta$. Then, we have:
\begin{equation}
v_c=\frac{\kappa v_a\alpha^2(1+\beta)+\kappa v_a(1-\beta)-2\alpha\pi_a}{\kappa[\alpha^2(1+\beta)(H_1+H_2)+(\beta-1)(H_1-H_2)]}.
\end{equation}
Note that when $H_1=H_2$, corresponding to $\beta=1$, this result
gives the same scaling as for the simple rectangular shape.  

With $\alpha\gg1$, corresponding to a highly viscous cytoskeleton, we have
\begin{equation}
v_c\approx\frac{v_a}{H_1+H_2}-\frac{2\pi_a}{\kappa\alpha(1+\beta)(H_1+H_2)},
\end{equation}
which clearly shows that the  cell speed is scaling inversely with the average height $(H_1+H_2)/2$.

\section{Test of model predictions}

The above analysis indicates that the ratio of the height of stress band $\lambda$ and the cell height $H$ determines the cell speed. 
Thus, cells with equal ratio should have similar speeds. 
To test this explicitly, we simulated cells in chambers with
heights varying between $h=4 \mu m$ and $h=10 \mu m$, 
constraining the cell's height, while keeping the
ratio $\lambda/h=0.5$. 
Cells shapes for three different chamber heights are shown in 
Fig. \ref{pro_height}a while the 
cell speed and effective height as a function of chamber height are
shown in Fig. \ref{pro_height}b and c, respectively.
Clearly, the results from Fig. \ref{pro_height}b show that the speed of the
cell is independent of the chamber height,  consistent with 
our model prediction.

In addition, our derived expression predicts that if $\lambda=H$, corresponding to
an active stress region that spans the entire height of the cell, 
the cell speed should be independent of the chamber height.  To verify this, we performed simulations of confined cells with the active stress at the entire
cell front. To this end, we no longer constrain the stress to a 
narrow band and, instead, use $\sigma^{a}=-\phi\rho_a(1-\chi)\epsilon|\nabla\phi|^2\hat{\bm{n}}\hat{\bm{n}}$.
We introduce the factor of $1-\chi$ to prevent
protrusion in the region where the cell and substrate overlap,
something that is excluded from occurring in other models when the
band restricts protrusion. Resulting cell shapes for different chamber heights are
shown in Fig. \ref{whole_front}a. In Fig. \ref{whole_front}b, we plot the cell speed as a function of the chamber height and in the 
Fig. \ref{whole_front} we plot the effective height. As expected,  the cell speed changes little as the 
chamber height is varied, again consistent with our predictions.

\section{Oscillatory protrusions}
Results in the main text are for cells with constant active stress, resulting in constant cell shapes. Such constant shapes are
applicable to fish keratocytes, fast moving cells that
maintain their morphology \cite{Keretal08-s}.
Other cell types, however, including neutrophils and {\it Dictyostelium
discoideum} cells \cite{LevRap13-s}, move in a more time-dependent way, with repetitive and 
short-lived protrusions called pseudopods.  To determine the dependence of 
 cell speed on chamber height for these types of cells  we introduce an oscillatory modulation to the active stress: $\sigma^{a}=-\phi\rho_aG(\psi)\sin(2\pi t/T)\epsilon|\nabla\phi|^2\hat{\bm{n}}\hat{\bm{n}}$. Here,
   $T$ is the period of the oscillation cycle which can be varied. 
   Results from additional simulations show that the cell speed gets larger as 
   the substrate adhesion is  increased (Fig. \ref{osci_act}a). This dependence on adhesion was found to be largely independent of the period and is 
   similar to the one found for model cells with constant stress (Fig. 2b).  
   Also consistent with our results in the main text (Fig. 2c),  the effective height  is again inversely related to the adhesion strength.

\section{Experiments}
\subsection{Cell culture and preparation}
Wild type Dictyostelium discoideum (AX4) cells were transformed with a construct in which the regulatory region of actin 15 drives genes encoding a fusion of GFP to LimE ($\Delta$ coil LimE-GFP) and  a gene encoding a fusion of RFP to Coronin (LimE GFP/corA RFP)\cite{fuller2010-s}. Cells were  transformed with the plasmid pDM115 cAR1-RFP (Hygromycin resistance) to visualize the membrane. Cells were grown in a shaker,  containing 35.5g HL5 media ($^{\tiny{\textregistered}}$FORMEDIUM)/L of DI water\cite{sussman1987-s} in a shaker. 
%We used the culture having a doubling time of less than 8 hrs, because we found that the slower growing cells were less chemotactically active.  
When cells reached  their exponential phase ($1-2 \times 10^6$ cells/mL), they were harvested by centrifugation, washed in $\text{KN}_2$/Ca buffer (14.6 mM $\text{KH}_2\text{PO}_4$, 5.4 mM $\text{Na}_2\text{HPO}_4$, 100 $\mu$M $\text{CaCl}_2$, pH 6.4), and resuspended in $\text{KN}_2$/Ca at $10^7$ cells/mL. The washed cells were developed for 5h with pulses of 50 nM cAMP added every 6 min. 
%After 5h of development, cells appeared to be polarized as the cells length to width ratio was larger than 3. 

\subsection{Microfluidic device}
The design of microfluidic device used in the study is similar to the design of the devices that were previously used 
to study gradient sensing in yeast\cite{paliwal2007-s} and chemotaxis in Dictyostelium\cite{skoge2010-s,Skoetal14-s}. The microfluidic device (Fig. \ref{chamber}) consists of a lithographically fabricated silicone (polydimethylsiloxane, PDMS, Sylgard 184) chip and a cover glass substrate (with either PDMS or hydrogel coating, see below), against which the chip is sealed using vacuum suction. To this end, the network of liquid-filled microfabricated microchannels of the chip, which are relatively narrow and either 100 or 10 $\mu$m  deep, is surrounded by a wide ($\sim$6 mm) and deep ($\sim$1 mm)  groove, serving as a vacuum cup. When the PDMS chip is placed on a substrate, the application of vacuum to the cup generates a pulling force that instantly seals the liquid-filled microchannels of the chip against the substrate. The application of vacuum also leads to controlled partial collapse of the microchannels, making it possible to reduce the depth of the 10 $\mu$m deep microchannels by $>5 \mu$m  by controlling the level of vacuum. 
The network of liquid-filled microchannels of the device
(Fig. \ref{chamber}) has a single outlet (out), two main inlets, for a $C_0=$100nM solution of cAMP (in 1) and for buffer (in 2), and an auxiliary inlet for cell loading (in c). The functional region of the device has two mirror-symmetric 100 $\mu$m deep, 500 $\mu$m wide flow-through channels (Fig. \ref{chamber}), which are connected to the two main inlets and are flanking 3 clusters of 10 $\mu$m deep gradient chambers. The flow through the device is driven by applying equal differential pressures of $\sim$2 kPa between the two main inlets and the outlet. The resulting mean flow velocity in the 500 $\mu$m wide flow-through channels is $\sim$200 $\mu$m/s. The gradient chambers are all 70 $\mu$m wide and each cluster has 15 identical chambers with equal lengths. The lengths $L$ of the gradient chambers in the upstream, middle, and downstream clusters are 360, 220, or 120 $\mu$m, respectively. There is practically no flow through the gradient chamber because of near zero pressure gradient along them, and the diffusion of cAMP from the flow-through channel perfused with the 100 nM solution to the flow-through channel perfused with buffer results in linear concentration profiles of cAMP with gradients of 0.28, 0.45, and 0.83 nM/$\mu$m, respectively. In different sets of experiments, the application of different levels of vacuum resulted in the effective depths of the gradient chambers of 10, 7, and 5 $\mu$m.

\subsection{Substrate preparation}
In our experiments, the microfluidic chips were sealed against cover glass substrates with two different types of coating: $\sim$10 $\mu$m thick layer of PDMS of the same type as the material of the chip and $\sim$3 $\mu$m thick layer of 30\%
 polyethylene glycol (PEG) gel. In the former case, the cover glass was a \#1.5 thickness 47 mm circle at the bottom of a 50 mm WillCo cell culture dish. A small amount ($\sim$0.2 mL) of PDMS pre-polymer (10:1 mixture of base and curing agent of Sylgard 184 by Dow Corning) was dispensed onto the cover glass.
  Spin-coating was made at 6000 rpm for 2 min, and PDMS was cured by overnight baking in a 60\si{\degree}C oven. In the latter case, the cover glass was \#2, 50x35 mm rectangle. The cover glass was cleaned with water and ethanol, dried, air-plasma treated for 10 s, and then  exposed to 3-(Trimethoxysilyl) propyl Methacrylate ($^{\tiny{\textregistered}}$Aldrich) vapor at 75\si{\degree}C for 30 min. A 30\% PEG pre-polymer solution was prepared by mixing PEG diacrylate (PEG-DA; avg Mn 900, $^{\tiny{\textregistered}}$Aldrich) with a 0.03\% aqueous solution VA086 (300 $\mu$g dissolved in 1000 $\mu$L of DI water) in a 3:7 ratio by volume. VA086 is iLine (365nm) sensitive UV photo-initiator
   that cross-links PEG-DA molecules (thus, converting a PEG-DA solution into a PEG gel)  by binding to the acrylate groups and also links PEG-DA chains to the acrylate groups on the glass surface. An $\sim$100$\mu$L drop of the solution was dispensed onto the center of the cover glass and squeezed to a thin layer by placing an untreated \#1.5 thickness, 30 mm diameter round cover glass on top, gently pushing this second cover glass with a pipette tip, and removing the excess solution with a wipe. The cross-linking of PEG-DA was done by exposing it to a total of 2.19  $J/cm^2$ of 365 nm UV (derived from 365nm UV LEDs; $\sim$365 $mW/cm^2$ for 60 sec). After the round cover glass was removed, the 50x35 mm cover glass had an $\sim$4 $\mu$m thick layer of covalently bonded PEG gel in the middle.
  
\subsection{Data acquisition and image analysis}
Differential interference contrast (DIC) images were taken of all gradient chambers on a spinning-disk confocal Zeiss Axio Observer inverted microscope using a 10x objective and a Roper Cascade QuantEM 512SC camera.  DIC images were captured every 15 s for 30 min and were used to calculate the speed of the cells. To obtain the shape of the cells, fluorescent images (488 nm and 561 nm excitation) were captured every 2 seconds with a 63X oil objective. To visualize the shape of the cells near the 
substrates,
z-stacks of confocal images were collected.

The centroids of all cells were tracked across the gradient chambers from 10X image sequences  using Slidebook 6 (Intelligent Imaging Innovations) software. Cells that moved more than 5 frames without encountering another cell were chosen for data analysis. 50 to 100 cell tracks were analyzed in each experiment. Velocity in the gradient direction, $V_x(t)$, was computed using data from frames 45 s apart with
  Matlab R2016a (The MathWorks, Natick, MA). We have verified 
  that cell 
 speeds were largely independent of their positions within the 
  gradient chambers. 
  Consequently,  the average speed was defined as the mean  speed of all  cells at least 30$\mu$m  from 
 the sides of the chamber adjacent to the flow-through channels at all recorded times. 

Cells outlines near the top (PDMS chip) and the bottom (substrate, 
PDMS or PEG) of the gradient chambers were obtained from confocal fluorescence images at 63X magnification with a custom-made Matlab code, as follows. 
After removing the average background intensity value,  images were binarized using a threshold that was dependent on the cell's maximum intensity. Matlab algorithms were then used  to dilate images, to fill possible holes, to erode 
images,  to smooth images, and to provide information (area and outlines) about the connected pixels of the binary image. Finally,  
using the resulting images, 
we computed the ratio between the cell contact area at the top and bottom of the chamber and averaged this ratio over three time points for each cell.  

\subsection{Statistics and reproducibility}
Each experiment was carried out four or five times on different days and the data were averaged for N=200-300 cells. 
Cell
speed was found to be approximately normally distributed and p-values were computed with the unpaired t-test. For the area size ratio, the data distribution was not normal, and the Wilcoxon rank-sum test was used to obtain the p-values. 
The variations of the cell speed with the gradient chamber height and the type of substrate coating (PDMS vs. PEG) followed the same trends in gradient chambers of different lengths, $L$ (cf. Fig. 4d and Fig. \ref{speed_length}).

%\bibliography{cell_motility}

\clearpage

%\begin{tabular}{l*{7}{c}r}
\begin{table}[ht]
	\begin{ruledtabular}
		\begin{tabular}{ccc}
			Parameter & Description & Value \\
			\hline
			$\gamma$ & Tension & 20 pN \\
			$\epsilon$ & Width of phase field  & 2 $\mu$m \\
			$A_0$ & Cell area size & 120 $\mu \text{m}^2$ \\
			$M_a$ & Cell area conservation strength & $20\;\text{pN}/\mu m$ \\
			$\Gamma$ & Phase field relaxation parameter & $0.4\;\mu m/s$ \\
			$\nu$ & Cell viscosity & $10^2\;\text{pN s}/\mu m$ \\
			$\xi_d$ & Damping coefficient & $0.05\;\text{Pa s}/\mu m$ \\
			$\xi_s$ & Substrate friction coefficient & $5\;\text{Pa s}/\mu m$ \\
			$\eta_a$ & Active protrusion coefficient & $10^3\;\text{pN}\;\mu m^2$ \\
			$\lambda$ & Width of active stress confinement & $2\;\mu m$ \\
			$\delta$ & Width of the substrate phase field & $1\;\mu m$ \\
			$g$ & Substrate repellent coefficient & $5\times10^3\;\text{pN}/\mu m$
			%\end{tabular}
		\end{tabular}
	\end{ruledtabular}
	\caption{ Model Parameters}
\end{table}

\clearpage

\begin{figure} \centering
	\includegraphics[width=.45\textwidth]{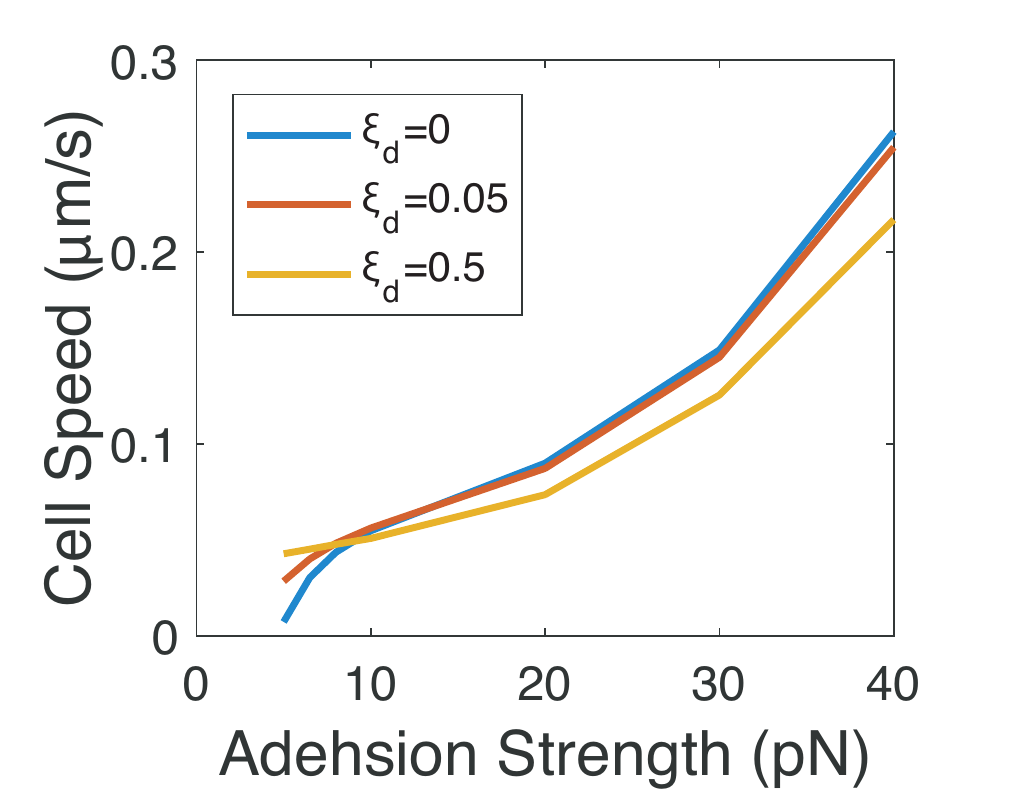}
	\caption{Cell speed as a function of adhesion strength for different values of 
		the drag coefficient  $\xi_d$ (in units of Pa s/$\mu$m). Cell speed changes little as $\xi_d$ is increased from 0 to 0.5.}
	\label{speed_xi0}
\end{figure} 

\begin{figure} 
	\centering
	\includegraphics[width=.45\textwidth]{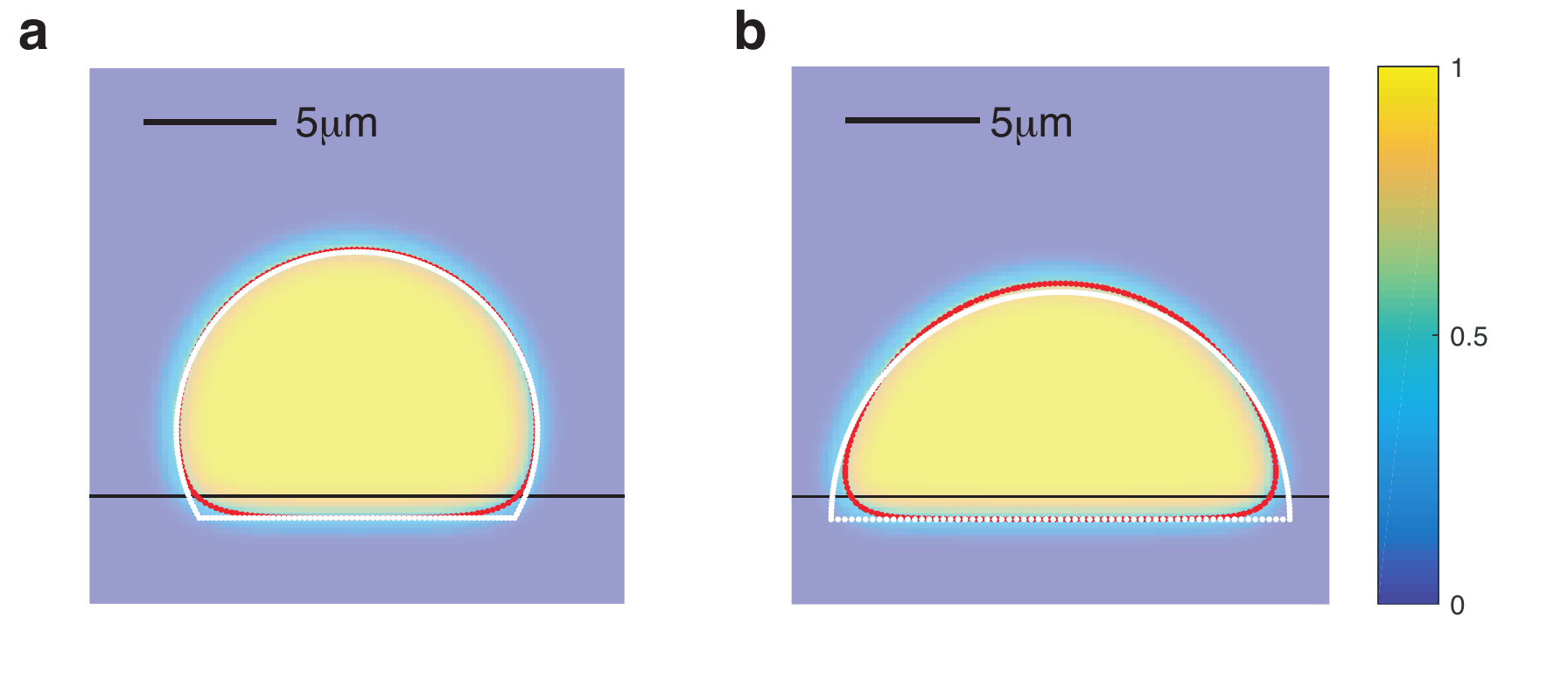}
	\caption{Simulation results of the phase-field method without active force (red line) compared to results obtained using Surface Evolver (white dots).
		The phase field is plotted using the indicated color scale. (a). Adhesion strength 10 pN. (b). Adhesion strength 20 pN.}
	\label{wetting}
\end{figure}

\begin{figure}\centering
	\includegraphics[width=.9\textwidth]{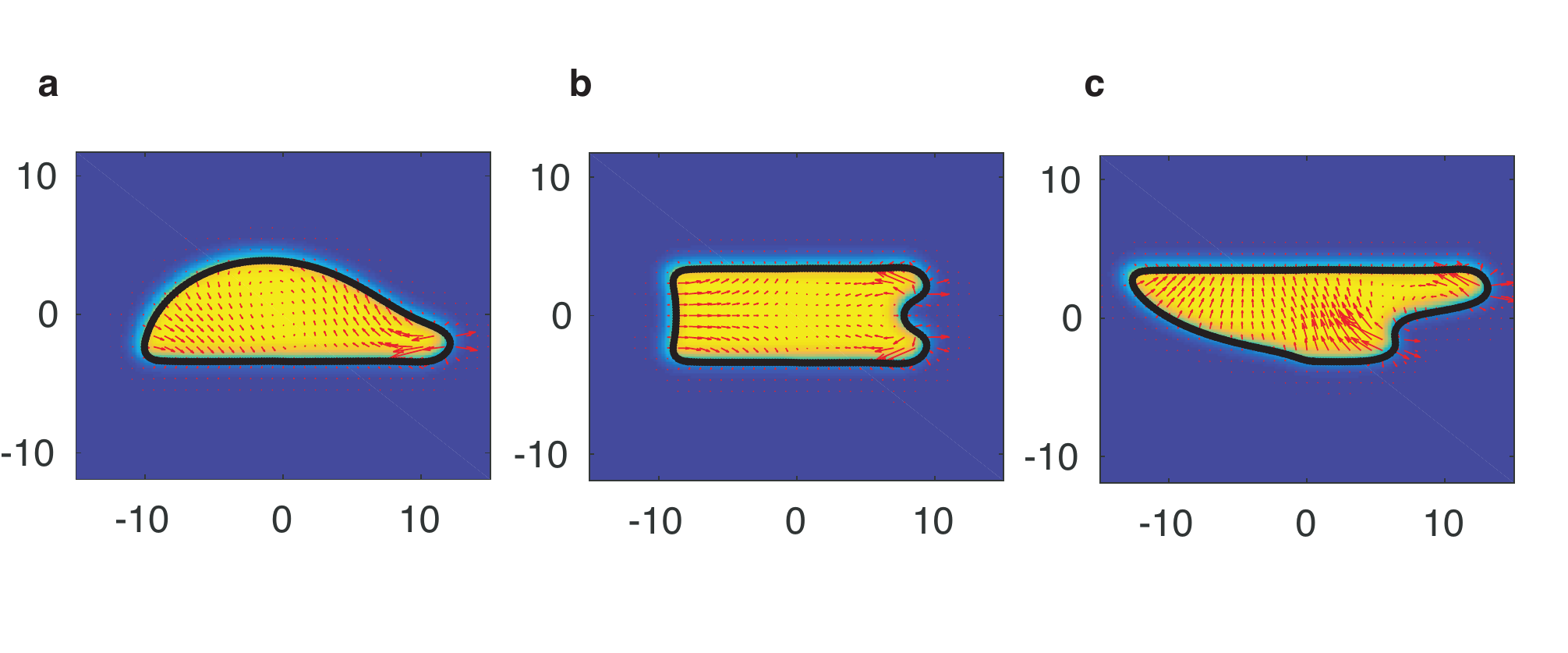}
	\caption{Numerical results showing the phase field using a color scale, the outline of the cell in black (defined as $\phi=1/2$), and the actin fluid velocity (multiplied with the phase field $\phi$) for a 
		cell moving on a  single substrate (a),  and confined in a
		channel with equal (b) and unequal substrate adhesion (c). Arrows indicate the direction of the velocity and the arrow length indicates the amplitude of the velocity.}
	\label{flow}
\end{figure}

\begin{figure}\centering
	\includegraphics[width=.4\textwidth]{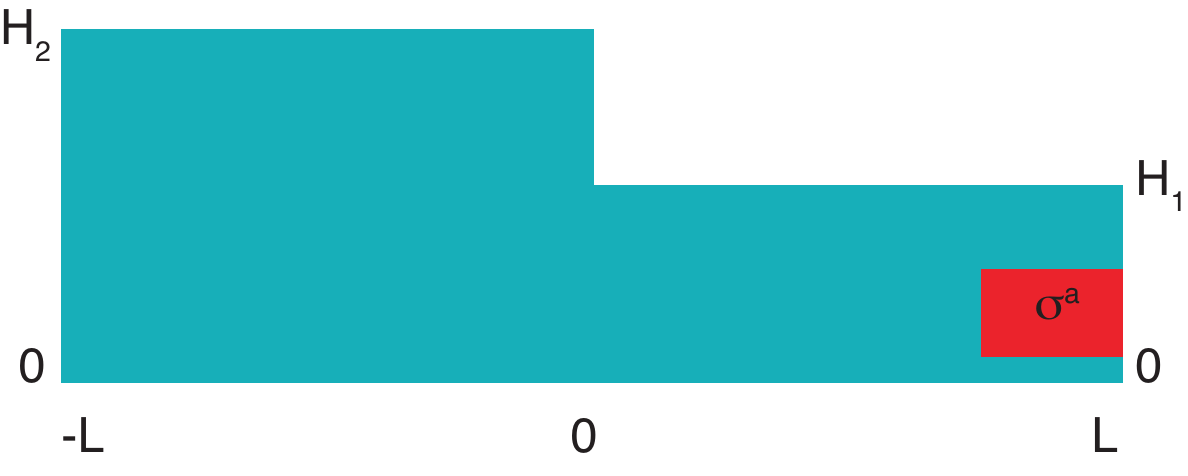}
	\caption{Schematic illustration of the non-deformable cell 
		considered here, consisting of two rectangles of unequal height.
		Active stress occurs in the right (front) rectangle.}
	\label{step}
\end{figure}

\begin{figure}\centering
	\includegraphics[width=.6\textwidth]{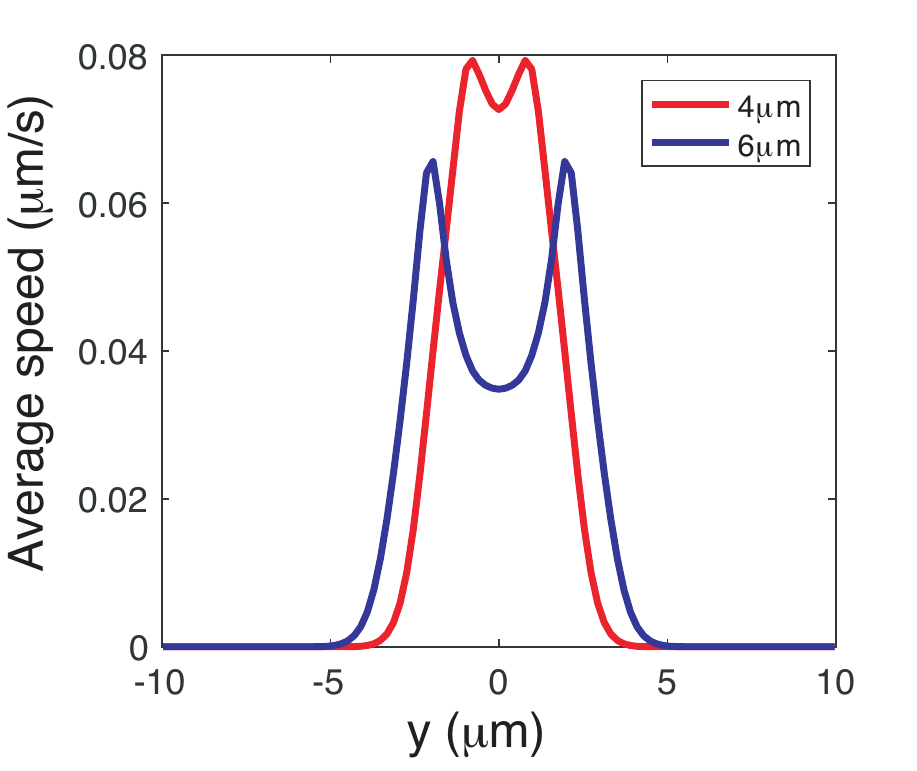}
	\caption{The average speed  along the y-direction, 
		defined as $1/L\int  |\phi u_x|dx$, for a cell in a confined chamber with a height of $h=4\mu$m (red line) and $h=6 \mu$m (blue line). The 
		vertical shear dissipation increases with increasing cell height.}
	\label{vel_prof}
\end{figure}

\begin{figure}\centering
	\includegraphics[width=.8\textwidth]{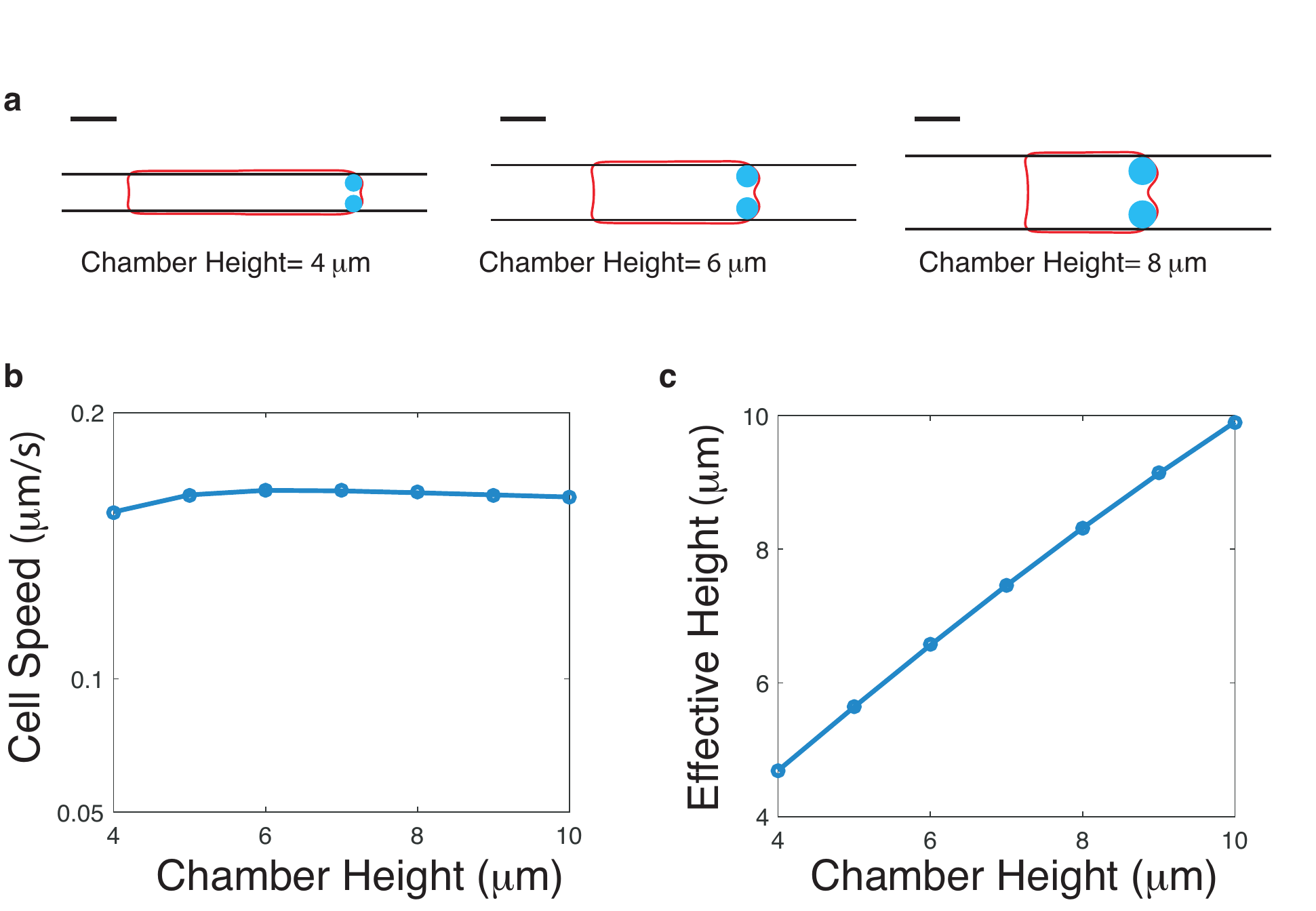}
	\caption{Simulation results of cells with a constant ratio 0.5 of the width of active stress band and the chamber height. The cyan dots schematically indicate the active stress sites.  (a) Cell shapes for different  chamber heights. Scalebar=5 $\mu m$ (b) Cell speed as a function of chamber height. (c) Effective height as a function of chamber height.}
	\label{pro_height}
\end{figure}

\begin{figure}\centering
	\includegraphics[width=.8\textwidth]{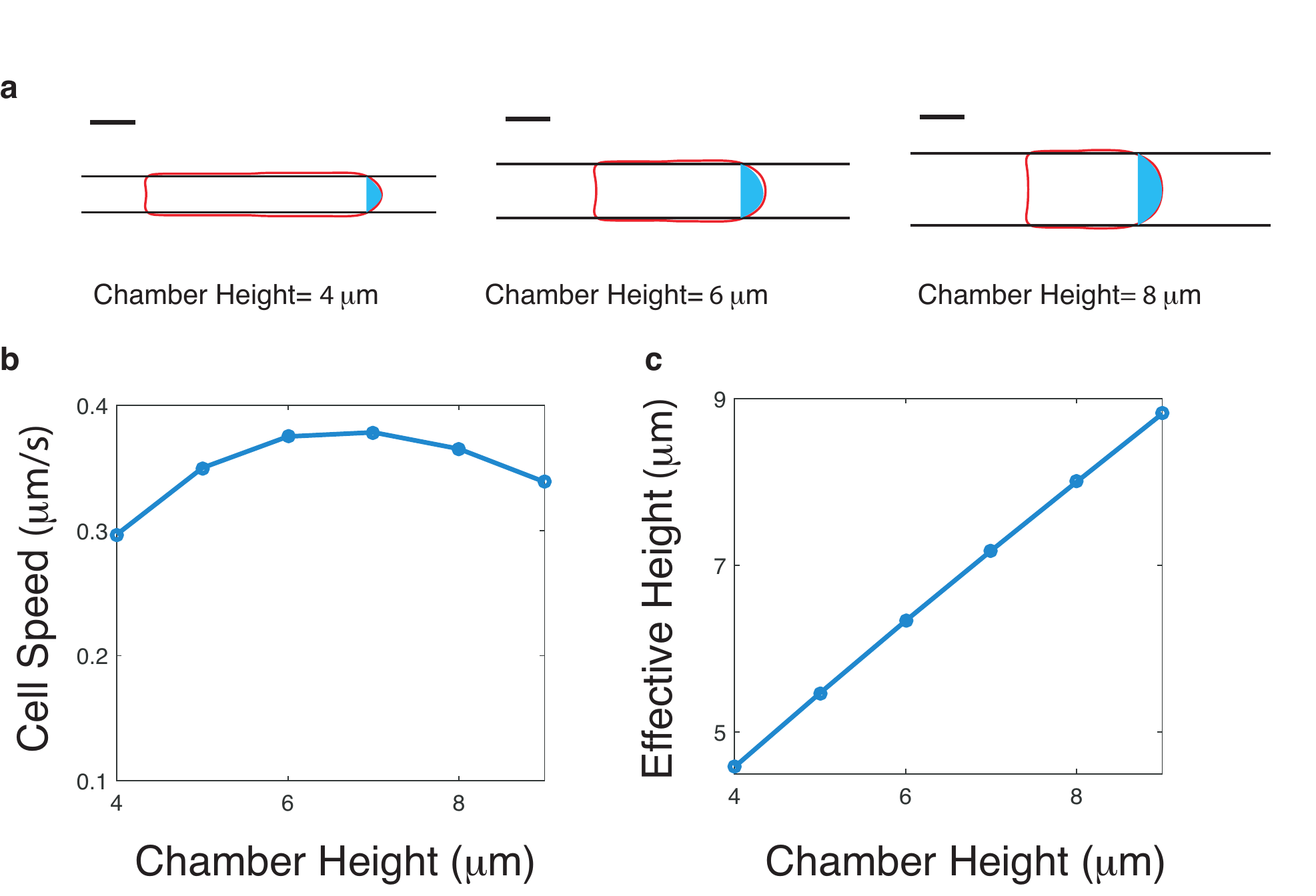}
	\caption{Simulation results of cells with active stress at the entire front, 
		as indicated. (a) Cell shapes for different  chamber heights. Scalebar=$5 \mu$m (b) Cell speed as a function of chamber height. (c) Effective height as a function of chamber height.}
	\label{whole_front}
\end{figure}

\begin{figure}\centering
	\includegraphics[width=.5\textwidth]{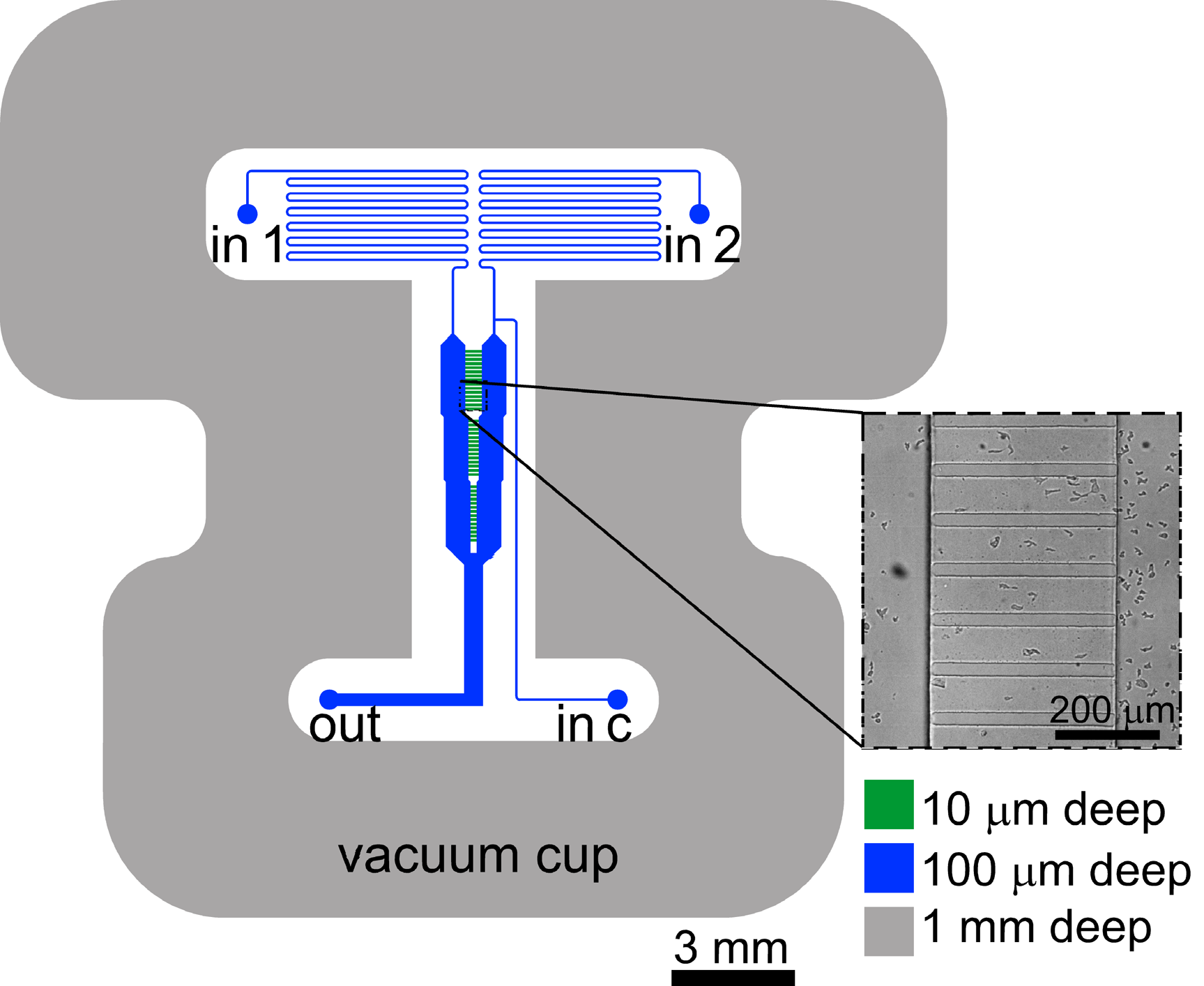}
	\caption{Design of the microfluidic device. The enlarged image is the experimental DIC view using a 10x objective showing  gradient chambers and flow chambers with cells.}
	\label{chamber}
\end{figure}

\begin{figure}\centering
	\includegraphics[width=.85\textwidth]{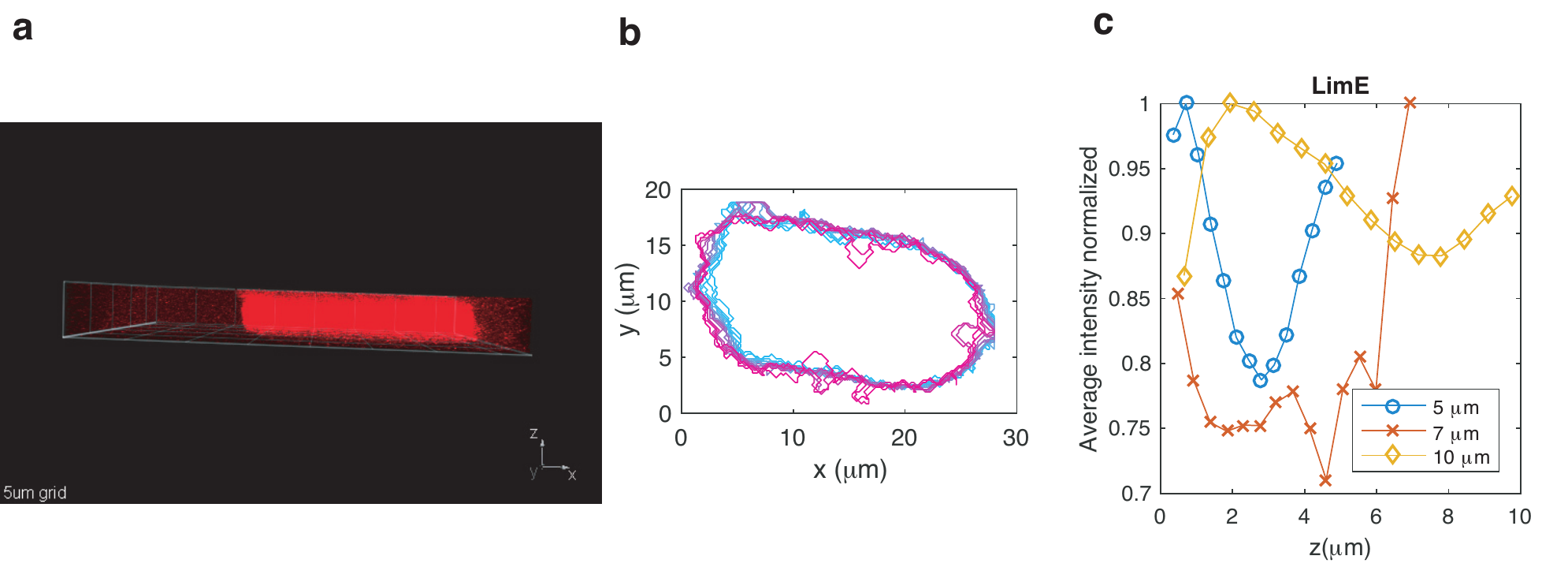}
	\caption{Experimentally obtained cell shapes and F-actin distribution for a cell moving in a 5 $\mu$m  high channel. (a) z-stack of a cell containing the  fluorescent membrane marker Car1-RFP. The cell extends from top to bottom PDMS substrate. (b)  Cell outlines for different z values ranging from 0 (magenta) to 5 $\mu$m (cyan). The outline is essentially identical for all z values. (c)  Average  fluorescence intensity (normalized) of LimE, an F-actin marker, for each confocal slice as a function of z for representative cells in channels with height of 5,7 and 10 $\mu$m. All cells examined (N=5) displayed a qualitatively similar pattern with increased intensity close to the substrates. } 
	\label{z-stack}
\end{figure}

\begin{figure}\centering
\includegraphics[width=.8\textwidth]{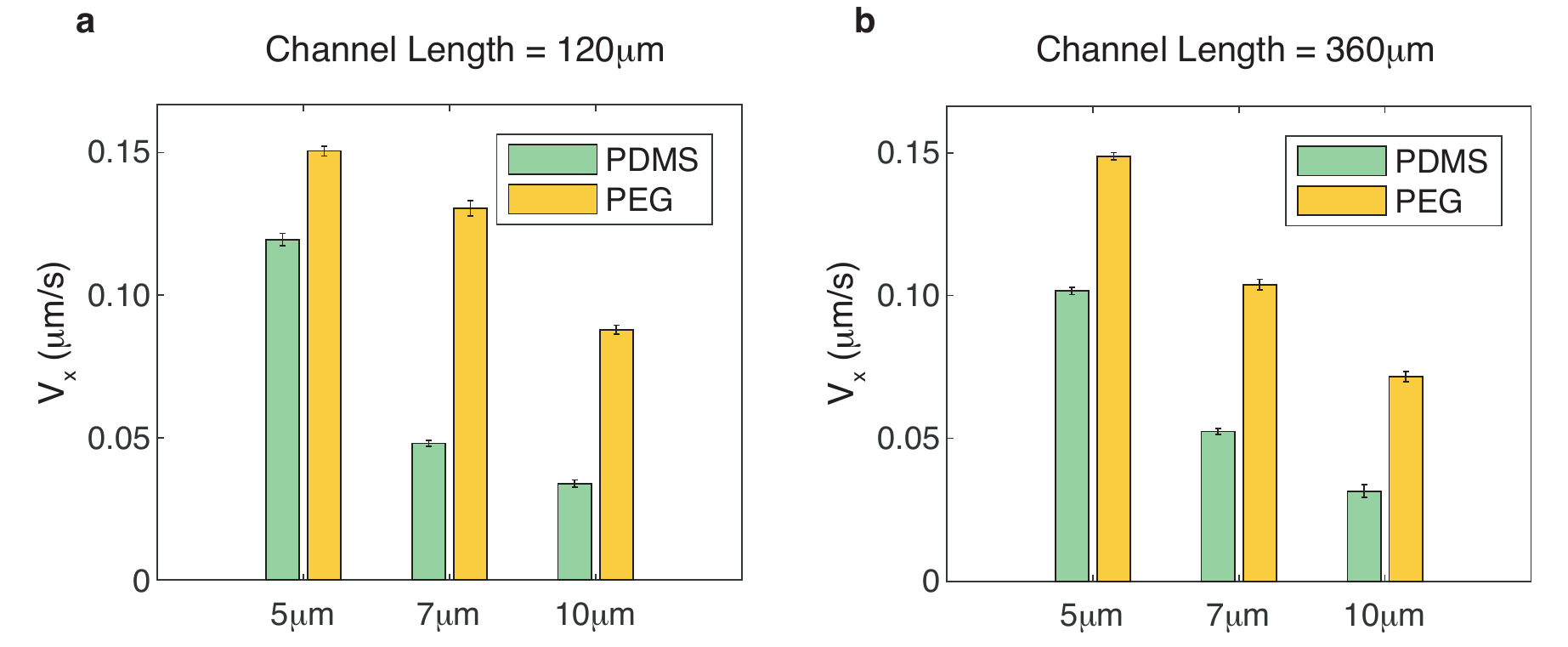}
\caption{Average cell speed for different chamber heights and 
	substrate composition for channel length $L=120\mu$m, corresponding to a gradient of 0.83 nM/$\mu$m, and $L=360  \mu$m, corresponding to a gradient of 0.28 nM/$\mu$m.}
\label{speed_length}
\end{figure}

\begin{figure}\centering 
	\includegraphics[width=.8\textwidth]{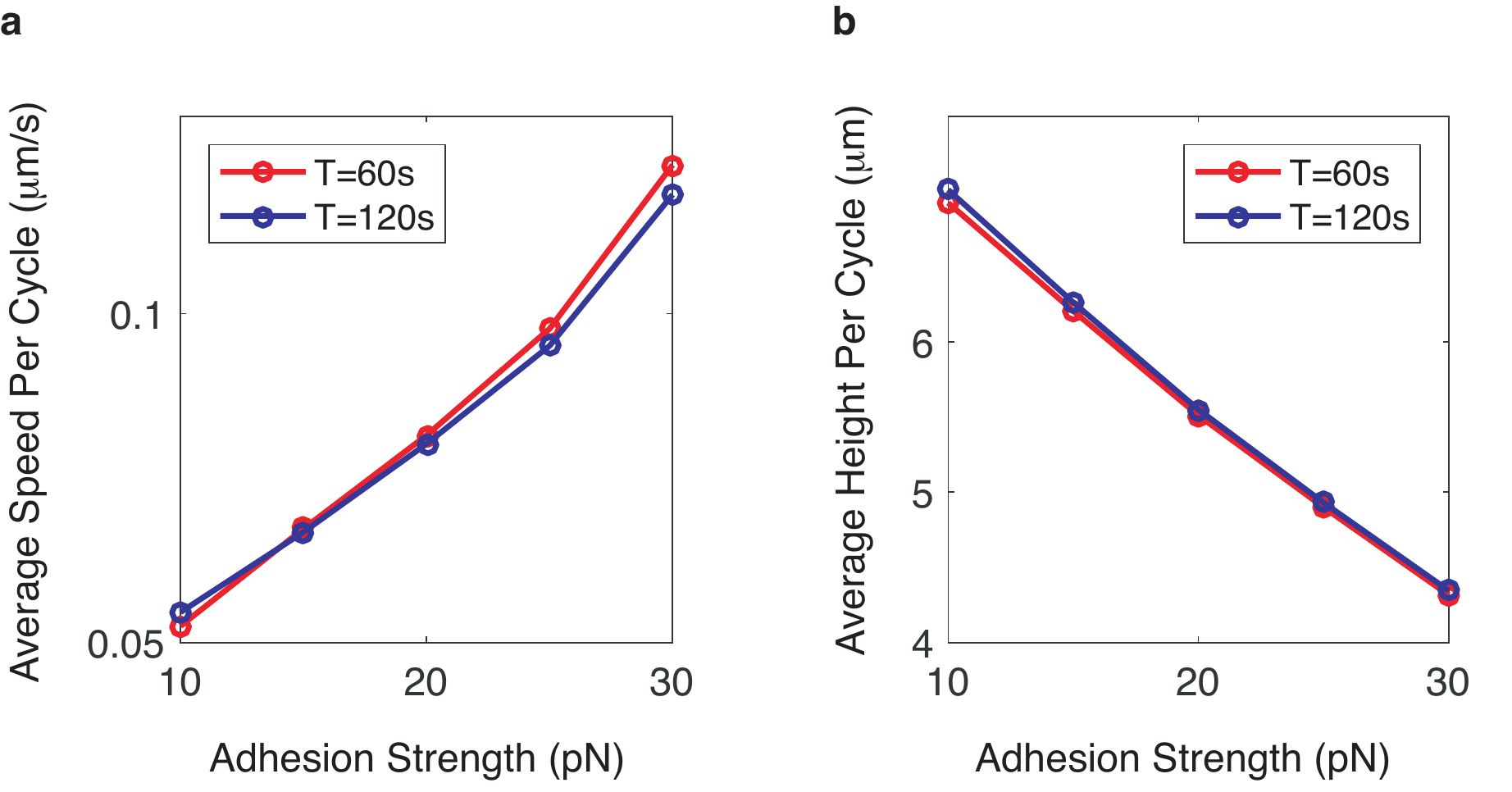}
	\caption{Cell speed and effective height for cells with oscillatory active stress. (a) Average cell speed, computed as moving distance divided by cycle time, and (b) effective height  as a function of substrate adhesion strength. Shown are the results for oscillatory stress cycles with two 
		different periods.}
	\label{osci_act}
\end{figure}

%

%\section*{Figures}

%%%REFERENCES%%%
%\bibliography{cell_motility} %You need to replace "rsc" on this line with the name of your .bib file

\end{document}